\documentstyle[11pt,aaspp4]{article}

\def\msol{{\cal M_{\odot}}}
\def\mbol{M_{\rm bol}}

\def\teff{{T_{\rm eff}}}

\def\lta{\mathrel{\hbox{\raise 0.6 ex \hbox{$<$}\kern
                   -1.8 ex\lower .5 ex\hbox{$\sim$}}}}
\def\gta{\mathrel{\hbox{\raise 0.6 ex \hbox{$>$}\kern
                   -1.7 ex\lower .5 ex\hbox{$\sim$}}}}
\def\br{\hbox{$B\!-\!R$}}		
\def\bi{\hbox{$B\!-\!I$}}		
\def\vi{\hbox{$V\!-\!I$}}

\slugcomment{Astrophysical Journal, accepted}

\lefthead{Bergbusch \& VandenBerg}
\righthead{$\alpha$-Element Enhanced Stellar Models. III. IPFs}

\begin{document}

\title{Models for Old, Metal-Poor Stars with Enhanced $\alpha$-Element
Abundances. III. Isochrones and Isochrone Population Functions}

\author{Peter A.~Bergbusch}
\affil{Dept.~of Physics, University of Regina, Regina, Saskatchewan, S4S~0A2,
       Canada\\Electronic mail: bergbush@phys.uregina.ca}

\and

\author{Don A.~VandenBerg} 
\affil{Dept.~of Physics \& Astronomy, University of Victoria, 
       P.O.~Box 3055, Victoria, B.C., V8W~3P6, Canada\\ 
       Electronic mail: davb@uvvm.uvic.ca}
 
\begin{abstract} 
An isochrone population function (IPF) gives the relative distribution
of stars along an isochrone. IPFs contain the information needed to
calculate both luminosity functions and color functions, and they
provide a straightforward way of generating synthetic stellar
populations. An improved algorithm for interpolating isochrones and
isochrone population functions, based on the scheme introduced by
Bergbusch \& VandenBerg (1992, \apjs, 81, 163), is described. Software
has been developed to permit such interpolations for any age
encompassed by an input grid of stellar evolutionary tracks.  Our first
application of this software is to the models presented in this series
of papers for 17 [Fe/H] values between $-2.31$ and and $-0.3$, with
three choices of [$\alpha$/Fe] at each iron abundance (specifically,
0.0, 0.3, and 0.6). [These models do not treat gravitational settling
or radiative acceleration processes, but otherwise they are based on
up-to-date physics. Additional grids will be added to this data base as
they are completed.] The computer programs (written in FORTRAN 77) and
the grids of evolutionary tracks which are presently available for
processing by these codes into isochrones and IPFs are freely available
to interested users.  In addition, we add to the evidence presented in
previous papers in this series in support of the $\teff$\ and color
scales of our models.  In particular, the temperatures derived by
Gratton et al.~(1996, A\&A, 314, 191) for local Population II subdwarfs
with accurate ({\it Hipparcos}) parallaxes are shown to be in excellent
agreement with those predicted for them, when the Gratton et al.~[Fe/H]
scale is also assumed.  Interestingly, the locus defined by local
subdwarfs and subgiants on the $(M_V,\,\log T_{\rm eff})$ plane and the
morphologies of globular cluster C-M diagrams are well matched by the
present models, despite the neglect of diffusion --- which suggests
that some other process(es) must be at play to limit the expected
effects of gravitational settling on predicted temperatures.  The three
field halo subgiants in our sample all appear to have ages $\gta 15$
Gyr, which is favored for the Galaxy's most metal-poor globular
clusters (GCs) as well.  (The settling of helium and heavy elements in
the central regions of stars is expected to cause about a 10\%
reduction in these age estimates: this effect should persist even if
some process, such as turbulence at the base of the convective
envelope, counteracts diffusion in the surface layers.)  Furthermore,
our isochrones accurately reproduce the Da Costa \& Armandroff (1990,
AJ, 100, 162) red-giant branch fiducials for M$\,$15, NGC$\,$6752,
NGC$\,$1851, and 47 Tuc on the [$M_I,\,(V-I)_0$]-diagram.  However, our
models fail to predict the observed luminosities of the red-giant bump
by $\approx 0.25$ mag: this could be an indication that there is some
amount of inward overshooting of convective envelopes in red giants.
For consistency reasons, the Zinn \& West (1984, ApJS, 55, 45)
metallicities for intermediate metal-poor GCs ($-1.8 \gta$ [Fe/H] $\gta
-1.1$) seem to be preferred over recent spectroscopic results (based on
the brightest cluster giants), suggesting that there is an
inconsistency between current subdwarf and GC [Fe/H] scales.

\end{abstract} 

 
\section{Introduction} 
Most metal-poor stars have relatively high abundances of the so-called
``$\alpha$-elements'' (O, Ne, Mg, Si, S, Ar, Ca, and Ti) --- which is
to say that the number abundance ratios O/Fe, Ne/Fe, etc., are higher
than one finds in the Sun (see, e.g., the reviews by \cite{wst89},
\cite{kra94}, and \cite{mcw97}).  For instance, \cite{car96} concluded
from his review of the available observational data that globular
clusters (GCs) with [Fe/H] $\lta -0.6$ are overabundant in the
$\alpha$-elements by a factor of about two (i.e., [$\alpha$/Fe]
$\approx +0.3$ in the standard logarithmic notation).  Since then, some
GCs have been discovered to have solar proportions of these
elements (see \cite{bwz97}).  Moreover, recent field star studies
(e.g., \cite{ns97}) have revealed a significant number of halo stars
with [$\alpha$/Fe] $\approx 0.0$, though the majority of such stars
show the same level of enhancement as is typically found in GCs.  Given
that the abundances of the individual $\alpha$-elements appear to scale
with the overall $\alpha$-element abundance,\footnote{It is our
impression from the scientific literature that the mean [$\alpha$/Fe]
value generally represents the abundance ratios determined for the {\it
individual} $\alpha$-elements to within 0.1 dex.  Oxygen may turn out
to be an exception to this rule, but current estimates of the abundance
of this element have large error bars associated with them because they
depend sensitively on which spectral features are analyzed:
much higher O/Fe ratios are found from the ultraviolet OH bands
(\cite{igr98}, \cite{bkdv99}) than from the $\lambda\,630.0,
\lambda\,636.4$ nm [O I] lines (\cite{fk99}).  In this investigation we
assume that all of the $\alpha$-elements, including oxygen, vary
together.  The possibility that [O/Fe] $\sim 1$ in extremely
metal-deficient stars, and the consequences of such high oxygen
abundances for synthetic C-M diagrams and for turnoff luminosity versus
age relations, is considered by \cite{vb01}.} the chemical composition
of a globular cluster or Population II star is often described in terms
of the two quantities [Fe/H] and [$\alpha$/Fe].

     It goes without saying that the models which are used to interpret
stellar data should be computed for the run of chemical abundances
that is actually observed --- and given that there are star-to-star
differences in the abundances of the $\alpha$-elements at a fixed iron
content, it is important to treat such variations explicitly.
Accordingly, a new grid of stellar evolutionary tracks has been
computed by \cite{vsr00} for 17 [Fe/H] values between $-2.3$ and $-0.3$
assuming, in each case, [$\alpha$/Fe] $=0.0$, 0.3, and 0.6.  These
calculations have also taken into account the most important non-ideal
effects in the equation of state as well as recent improvements to
opacity data and nuclear reaction rates (though gravitational settling and radiative acceleration
processes are not treated).  Moreover, as fully discussed
in Paper I, the properties of local sudwarfs, the loci for GC red
giants on the ($M_{\rm bol},\,\log T_{\rm eff}$)--plane as inferred
from infrared photometry, and the luminosities of RR Lyrae variables as
deduced from Baade-Wesselink, statistical parallax, and trignometric
parallax studies are reproduced quite well by the models.  Reference
should be made to that study for a full description of how variations
in [$\alpha$/Fe] affect the tracks and zero-age horizontal-branches,
both at a fixed [Fe/H] and as a function of [Fe/H].

     In Paper II, \cite{van00} examined the implications of the new
model grids for the ages of GCs and of field halo stars.  The main
results of that study are (1) the most metal-deficient clusters {\it
and} field subgiants have ages $\gta 14$ Gyr, (2) ages decrease with
increasing [Fe/H] (by 2--3 Gyr between [Fe/H] $\sim -2.3$ and $\sim
-1.3$) (3) the dispersion in age at a fixed metallicity appears to be
small (especially at [Fe/H] $\lta -1.3$), and (4) there is no more than
a weak variation of age with Galactocentric distance, if that.
However, the photometric data for many of the clusters are still not as
good as they need to be, relative age estimates based on the horizontal
method, as described by \cite{vbs90}, sometimes seem to be
inconsistent with those derived from the $\Delta V_{\rm TO}^{\rm HB}$
technique (suggesting that something besides, or in addition to, age is
varying), and the uncertainties associated with most reddening,
metallicity, and distance determinations remain large.  Consequently,
the conclusions that were reached must still be considered
tentative even though they seem well-supported by the observations {\it
as they presently exist}.  (Many age-related issues, such as how best
to intercompare synthetic and observed C-M diagrams for the determination
of both absolute and relative ages, are discussed in Paper II.)

     The primary purpose of the present investigation is to make the 
$\alpha$-element enhanced isochrones and what we have called ``isochrone
population functions'' (IPFs) readily available to the astronomical community.
An IPF is derived from the slope of the predicted mass--{\it distance} relation
in ${{\cal M}_\odot}$/mag, where the distance along an isochrone, on any
color--magnitude plane for which good color--$T_{\rm eff}$ relations are
available, is defined with respect to some arbitrary, well-defined point.  In
conjunction with an assumed mass spectrum, IPFs thus provide the
means to generate synthetic stellar populations in a very straightforward way
or to calculate luminosity functions (LFs) or color functions (CFs).  The
value and importance of LFs have long been appreciated (see, e.g., \cite{rf88}),
but it is only recently that \cite{bv97} have pointed out that the distibution
in color of stars along the subgiant branch has the potential to provide an
age constraint that is not only completely distance-independent, but is also
nearly metallicity-independent.  To obtain meaningful results from observed CFs,
superb photometry and very large subgiant samples are needed, but it should be
possible to meet these requirements in the case of the nearest and most massive
of the Galaxy's GCs.  Certainly, all aspects of observed C-M diagrams should be
fully investigated in order to achieve the best possible understanding of
stellar systems.

     In the following section, an improved algorithm for interpolating in
evolutionary sequences for isochrones and IPFs is described: this is based in
large part on the scheme introduced by \cite{bv92}.  Section 3 outlines how to
execute the software that has been developed to obtain isochrones, IPFs, LFs,
and CFs, and provides a fairly detailed description of the output of these
codes.  In \S 4, some additional discussion beyond that contained in Papers I
and II is presented on how well the present computations appear to satisfy
observational constraints.  Brief concluding remarks are given in \S 5.  
  
\section{The Interpolation Algorithm}

The mathematical formalism for the derivation of accurate IPFs
parallels that for the derivation of luminosity functions presented in
BV92, except that the distance along an isochrone is used as the
observable coordinate instead of luminosity.  In the $\log
L$--$\log\teff$ plane, the distance along an isochrone is defined by
$$d{\cal D} = \left[c_1\,(d\,\log L)^2 + c_2\,(d\,\log\teff)^2\right]^{1/2},
\eqno(1)$$ where $c_1$ and $c_2$ are chosen to stretch the SGB region
and to obtain a convenient range of distances.  (We have chosen
$c_1=1.25$ and $c_2 = 10.0$, and we arbitrarily define ${\cal D} =0$ at
the lowest mass point on the isochrone.) The slope of the mass--distance 
relation is then obtained from$${d{\cal D}\over d{\cal M}} =
\left[c_1\,\left({\partial\log L\over\partial{\cal M}}\right)^2 +\,
c_2\,\left({\partial\log\teff\over \partial{\cal
M}}\right)^2\right]^{1/2}.\eqno(2)$$ Consequently, equation (6) from
BV92 is recast as $$\phi({\cal D}) = N({\cal M}) \left({d{\cal D}\over
d{\cal M}}\bigg\vert_t\right)^{-1},\eqno(3)$$ where $N({\cal M})$ is
the number of stars with mass ${\cal M}$, and $\phi({\cal D})$ is the
differential isochrone population function.  (Note that the second of
these two equations becomes the familiar definition of a differential
luminosity function if ${\cal D}$ is replaced by $L$ --- see BV92.) The
evaluation of the partial derivatives in equation (2) is performed as
described in BV92.

In the observer's plane, the distance is calculated with respect to a
well-defined, but arbitrary, point on the isochrone. For this paper, a
point on the subgiant branch (SGB) 0.05~mag redward of the main-sequence
turnoff (regardless of color index),
has been adopted as the distance zero-point
with the idea that it should be relatively easy to identify in observed
CMDs. A convenient definition of distance is $$dD = \left[dM_i^2 +
16\,dC_{ik}^2\right]^{1/2}, \eqno(4)$$ where $M_i$ is the absolute
magnitude in the $i^{\rm th}$ passband and $C_{ik}$ is the color index
for passbands $i$ and $k$. However, the transformation of equation (3)
from the theoretical plane to the observer's plane involves the
derivatives of both the color transformations and the bolometric
corrections. Rather than attempt to evaluate these numerically, it is
easier to construct a $\cal D$--$D$ calibration which can be used to
translate bin limits from the observer's plane to the theoretical
plane, and then to perform integrations across bins on the theoretical
plane.

\subsection{Equivalent Evolutionary Phases}

Isochrones, IPFs, LFs, and CFs are interpolated from grids of
evolutionary tracks for stars of the same initial chemical composition
but different masses. Consequently, the
most fundamental interpolation is the one in which the mass is derived
from the age--mass relation for a given evolutionary phase. The obvious
advantage of an interpolation scheme in which mass is a monotonic
function of age is that the mass is uniquely determined once the age
and evolutionary phase are specified. The corresponding luminosities,
effective temperatures, and their temporal derivatives then can be
interpolated with respect to mass.

The basic approach to setting up the interpolation scheme remains as it
was described in BV92, with up to seven primary ``equivalent evolutionary
phase" (EEP) points (see, in particular, their Fig.~2) located on each 
track between the zero-age main sequence and the tip of the RGB.  The
definitions of two of these points have been revised slightly.  The
first primary EEP is now taken to be the point at which the central
hydrogen content falls to $X_i - 0.0022$, where $X_i$ is the adopted
H abundance in the initial fully convective model on the Hayashi line.
This corresponds quite well with the zero-age main-sequence location of
the model on the H-R diagram.

\begin{figure}
\figurenum{1}
\plotfiddle{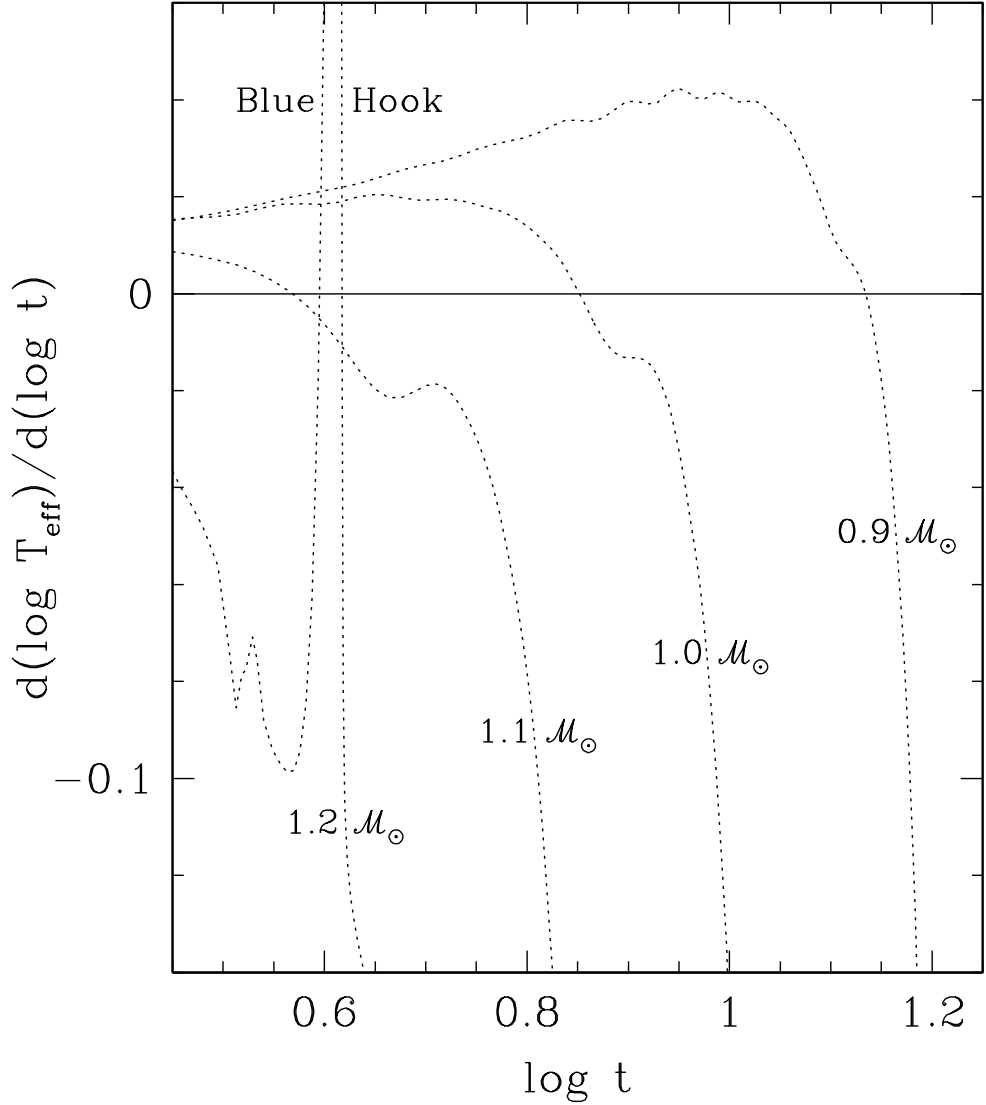}{6.5in}{0}{80}{80}{-252}{-72}
\caption{The derivative of $\teff$ with
respect to time in the region of the turnoff is plotted for selected
evolutionary tracks from a grid of models for [Fe/H] $=0.0$. The blue
hook in the $1.2\,{\cal M}_\odot$ track, which manifests itself as a
spike in this derivative, occurs at an age which is somewhat greater
than that of the turnoff point [where $d(\log\teff)/d(\log t) = 0.0$]
in the $1.1\,{\cal M}_\odot$ track.  To ensure that the age--mass
relation remains monotonic, the small dip after the turnoff point
in the 1.1 and $1.0\,{\cal M}_\odot$ tracks is used to define the
second primary EEP.\label{f1}}
\end{figure}

The more noteworthy change resulted from the discovery that the
age--mass relation defined by the second primary EEPs can be
non-monotonic just where the blue hook begins to manifest itself
(at $\approx 1.15 {{\cal M}_\odot}$), if
the amount of convective overshoot is assumed to increase with
increasing mass.  (Depending on whether the star has a radiative or
convective core during the main-sequence phase, the second primary EEP
corresponds to either the turnoff point or the minimum $T_{\rm eff}$
that is reached just prior to the blue hook, respectively.)  It is
however, relatively easy to avoid this difficulty.  As shown in
Figure~\ref{f1}, the competition between the $p$-$p$ chain and the CNO
cycle produces a small dip in the derivative $d\log \teff/d\log t$ for
the lower mass tracks that develops into the blue hook at higher mass.
When this feature appears {\it after} the turnoff point, it is used
to define the second primary EEP.\footnote{Minor oscillations in
the temperature derivative sometimes occur (e.g., those preceding the
turnoff point in the $0.9\,{\cal M_{\odot}}$ track in Fig.~1).
Perturbations to the model arising from the procedures used to
optimize the mesh point distribution or to calculate the atmospheric 
boundary conditions are the most likely causes of this behavior.
However, the amplitude of such variations is clearly very small and
of little consequence for computed IPFs.}
%The other three primary EEPs  occur at the local extrema
%in the temperature derivative at the Hertzsprung gap, the base of the
%red-giant branch (RGB), and the evolutionary pause on the giant
%branch, respectively.

Over the range 0.5--1.8~${\cal M}_{\odot}$ in stellar mass and for
metallicities $-2.31 \le\hbox{[Fe/H]}\le +0.12$ that we have
investigated, the age--mass relations for the primary EEPs remain
monotonic.\footnote{Extensions of the present calculations to higher
mass and [Fe/H] will be the subject of a forthcoming paper by
D.A.~VandenBerg, P.A.~Bergbusch, and P.D.~Dowler (in preparation).}
Thus, it is possible to maintain the monotonic behaviour in the
secondary EEP relations that are obtained by dividing each
evolutionary phase interval, bracketed by a pair of primary EEPs,
into an equal number of regularly placed sub-intervals. For the
evolutionary tracks of low-mass stars which achieve ages of 30 Gyr
before reaching the turnoff point, the placement of the last secondary
EEP is determined by matching it with the secondary EEP having the
same central hydrogen content on the track of the next higher mass.

\subsection{The Akima Spline}

In BV92, linear interpolation was found to be acceptable for all of the
EEP relations, except for the evaluation of derivatives along the
direction of interpolation. In the absence of a better solution at the
time, these derivatives were evaluated from a cubic polynomial fitted
through the EEP points --- in other words, the derivatives were
obtained from smoothed versions of the EEP relations. However, for this
paper, the interpolations and evaluations of derivatives were performed
with the Akima spline (Akima, 1970), for complete internal
consistency.  Among the various spline routines tested, the Akima spline
was found to be the best for tracing the subtle morphology of the EEP
relations, as it is particularly good at avoiding the typical spline
overshooting that occurs when the behaviour of the spline points is
non-polynomic.  In addition, as one may infer from
Fig.~6 of BV92, linear interpolation tends to produce the onset of
the convective hook in the isochrones at ages slightly greater than
those predicted by the evolutionary sequences.  Akima spline
interpolation reduces this discrepancy over the entire range of
metallicities that we have investigated.

\begin{figure}
\figurenum{2}
\plotfiddle{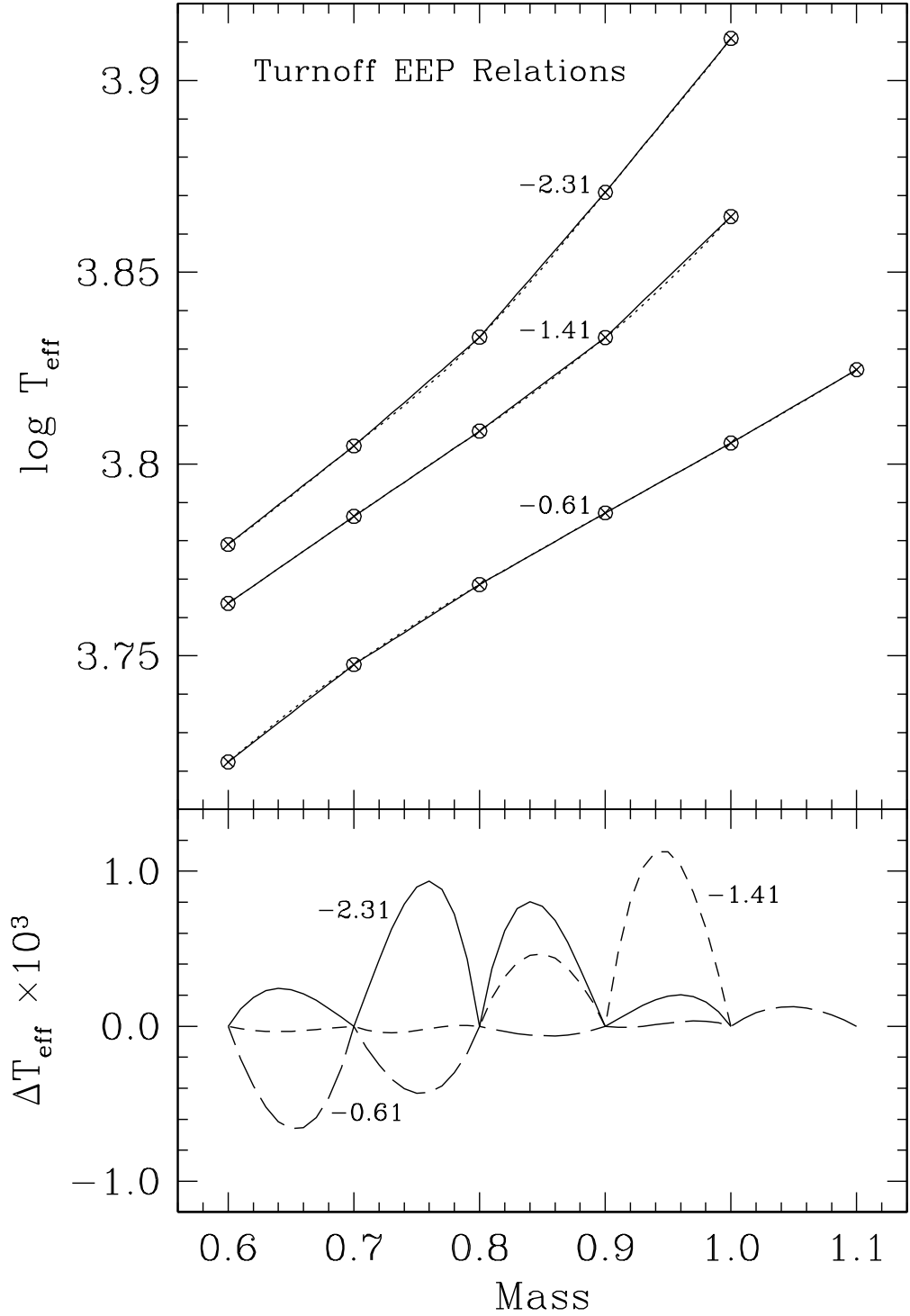}{7.75in}{0}{80}{80}{-252}{-24}
\caption{In the upper panel, the $\teff$ EEP relations at the
turnoff (the second primary EEP) are plotted for the metallicities
indicated from the [$\alpha$/Fe]$=0.0$ grid of models. For each
metallicity, the Akima spline fit is plotted as a {\it dotted curve};
linear interpolation is indicated by the {\it solid curve}. Differences
between the relations, in the sense $\rm [linear]-[Akima]$ are plotted
in the lower panel; the [Fe/H]$=-2.31$ differences are indicated by the
{\it solid curves}, those for [Fe/H]$=-1.41$ by the {\it short-dashed
curves}, and those for [Fe/H]$=-0.61$ by {\it long-dashed
curves}.\label{f2}}
\end{figure}

The differences between isochrones derived via Akima spline interpolation
and those derived using linear interpolation can be readily appreciated
by comparing the EEP relations that give the turnoff effective
temperatures and luminosities as a function of mass.  The
$\teff$~EEP relations, plotted in upper panel of Figure~\ref{f2},
reveal small, but detectable, differences between linear interpolation
and Akima spline interpolation at the turnoff, even for low mass stars.
Moreover, the largest differences always occur when
there is an abrupt change in the behaviour of the EEP relation. For
example, in the [Fe/H]$= -2.31$, [$\alpha$/Fe]$=0.0$ grid, isochrones
with ages 8 and 12 Gyr, interpolated by either method, are virtually
identical at the turnoff because they have turnoff masses near 0.9 and
0.8 solar masses, respectively. However, the turnoff masses for ages
10, 14, and 16 Gyr, lie near $0.85\msol$, $0.76\msol$, and $0.74\msol$,
respectively; i.e., approximately midway between the computed tracks.
The $\teff$ EEP relations shows a distinct change in slope at $0.8\msol$,
around which the spline deviates noticeably from the linear relations.
As can be seen from the lower panel of Fig.~2, the difference in
$\log\teff$ for a mass of $0.76\msol$ is $\approx 0.001$.  This means
that the temperature predicted by the Akima spline interpolation, for
the turnoff of a 14~Gyr isochrone, is $\approx 15~\rm K$ cooler than
that derived by linear interpolation.

\begin{figure}
\figurenum{3}
\plotfiddle{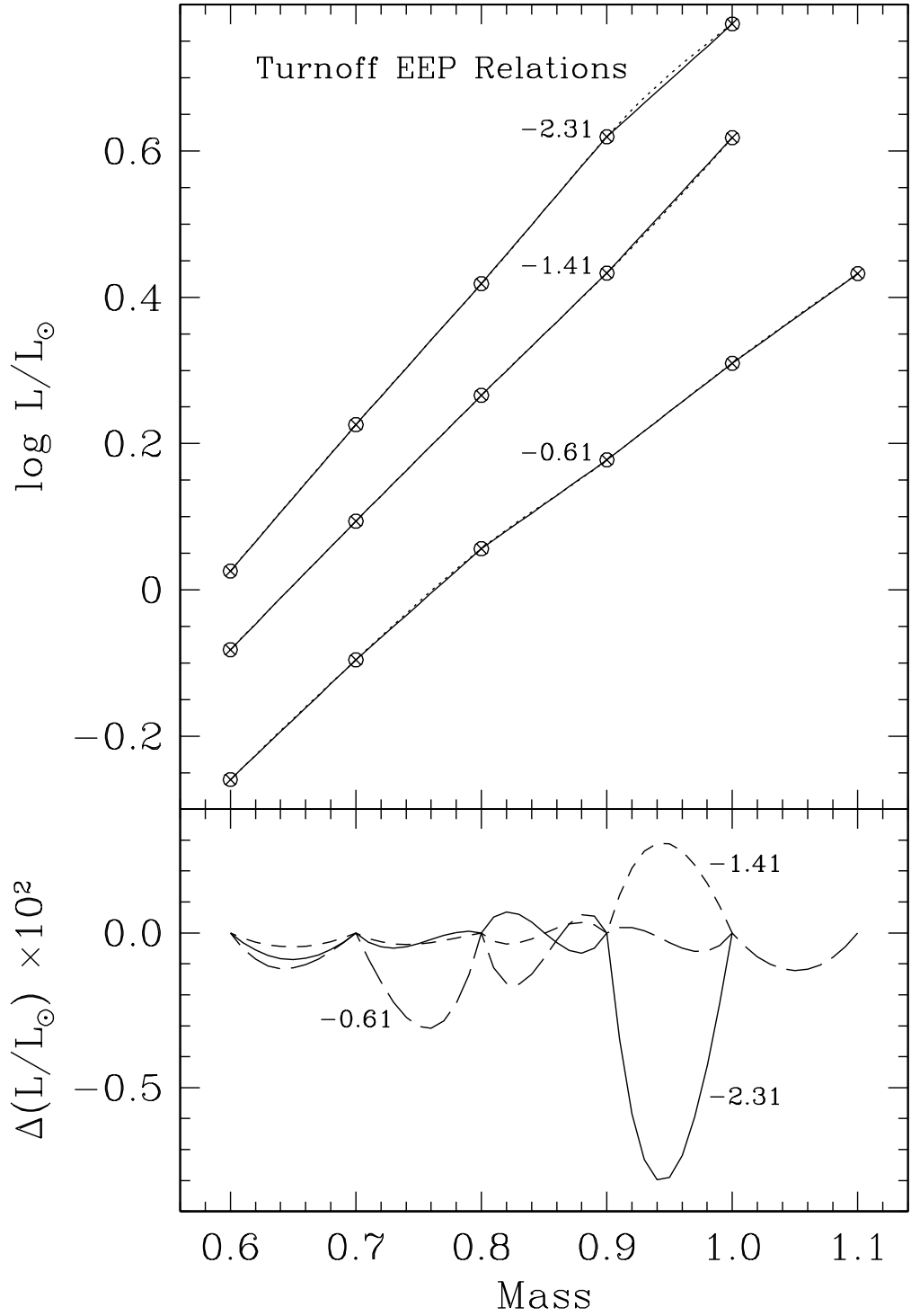}{8.0in}{0}{80}{80}{-252}{0}
\caption{Similar to Fig.~2, except the luminosity EEP
relations are plotted in the upper panel; the differences between the
linear and Akima spline versions of the relations are plotted in the
lower panel.\label{f3}}
\end{figure}

Similar observations can be made about the luminosity EEP relations,
plotted in the upper panel of Figure~\ref{f3}. The differences in
$\log L/L_{\odot}$ between the two interpolation methods, plotted in
the lower panel, reveal that turnoff luminosities are consistent to within
0.02~mag in $\mbol$ over the entire range of masses plotted, and to within
0.008~mag over the range of turnoff masses relevant to globular
cluster studies.

It should be emphasized that the points plotted in the EEP relations of
Figs.~2 and 3 are taken {\it directly} from the model tracks. In most
instances, they were identified automatically by software designed to
recognize the primary EEPs. Thus, they
directly reflect the behaviour of the models at the masses specified.
One problem that is exposed by the spline approach occurs as a result of
the fact that there are usually more tracks with zero-age main sequence
EEPs than turnoff EEPs, and more with turnoff EEPs than with EEPs on
the subgiant branch (because lower mass tracks are not extended past an
age of 30 Gyr).  Consequently, the nature of the spline fit to the
EEPs changes abruptly when the number of points used to construct the
fit changes. This sometimes manifests itself as a small discontinuity on an
isochrone generated via the Akima spline, that would not be evident on
one generated by linear interpolation.  (We emphasize that such
discontinuities are virtually undetectable; we have ensured this by
extending low mass tracks to the base of the RGB in the grids where
discontinuities appeared on isochrones near the turnoff.)

\begin{figure}
\figurenum{4}
\plotfiddle{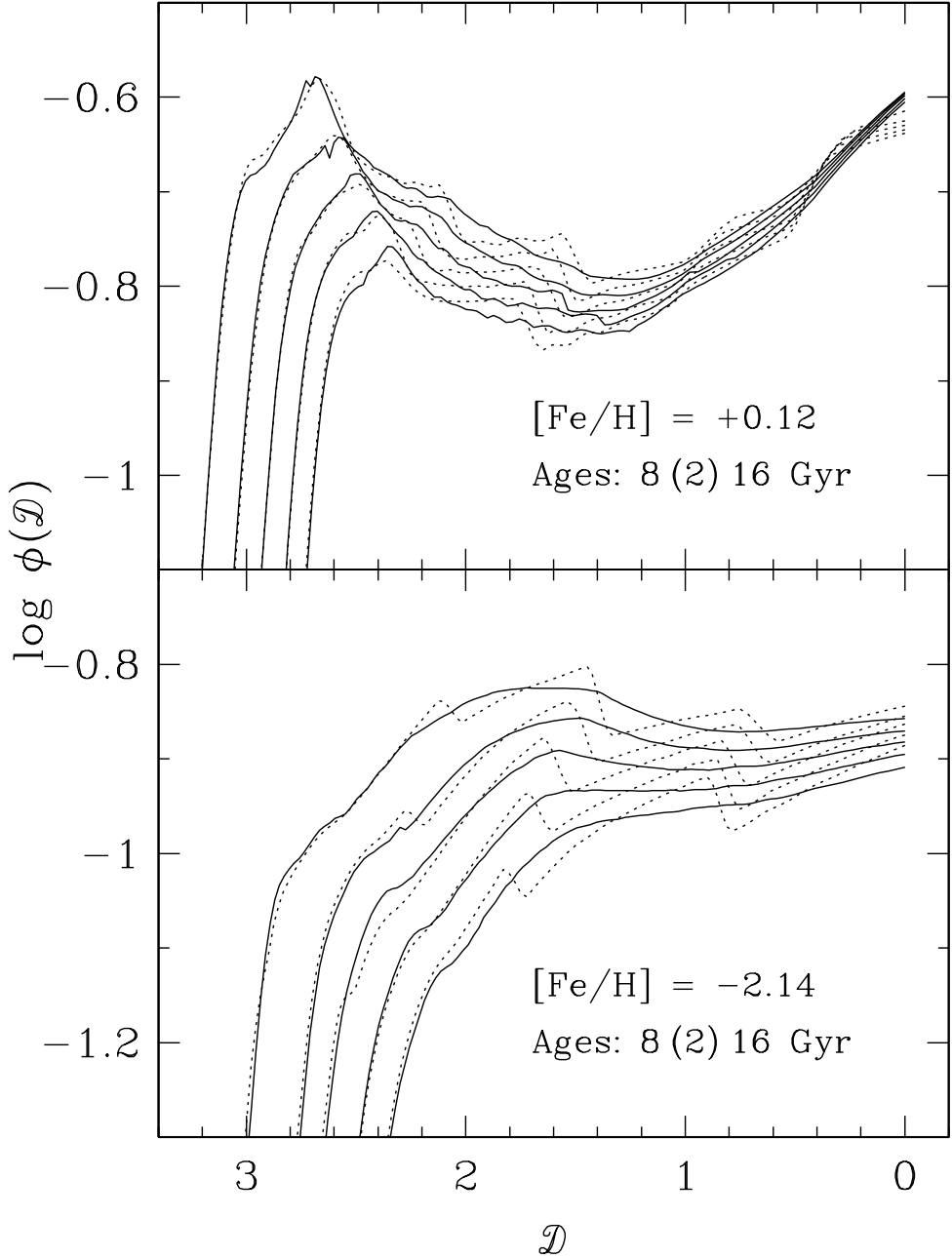}{7.5in}{0}{80}{80}{-252}{-54}
\caption{Differential IPFs on the theorist's plane from
the main sequence through the turnoff are plotted for the abundances
and ages indicated. Those obtained from Akima spline interpolation, are
plotted as {\it solid curves}; the corresponding IPFs obtained via
purely linear interpolation are plotted as the {\it dotted lines}. On
the theorist's plane, the zero-point of {\it distance along an isochrone} is
arbitrarily defined at the lowest mass point ($\approx\,0.5\msol$).\label{f4}}
\end{figure}

While the Akima spline provides an accurate means of
interpolating the points along an isochrone, its chief advantage over
linear interpolation is that derivatives can be evaluated directly at
each point. A comparison between differential IPFs on the
theorist's plane, derived from Akima spline interpolation via the
approach defined by equations (2) and (3), and IPFs calculated directly
from the tabulated masses and distances obtained via linear
interpolation, is shown in Figure~\ref{f4}. The discontinuities
evident on the main sequence portion of the linearly interpolated IPFs
occur wherever an isochrone crosses one of the evolutionary tracks.
They are caused by the abrupt changes in the slopes of the
interpolation relations at these points. Such discontinuities would
also be evident in linearly interpolated LFs and CFs; they did not
appear in the LFs presented in BV92 because the slopes of the EEP
relations were derived from cubic polynomial fits to the EEP points.

\subsection{Stability of the Models}

\begin{figure}
\figurenum{5}
\plotfiddle{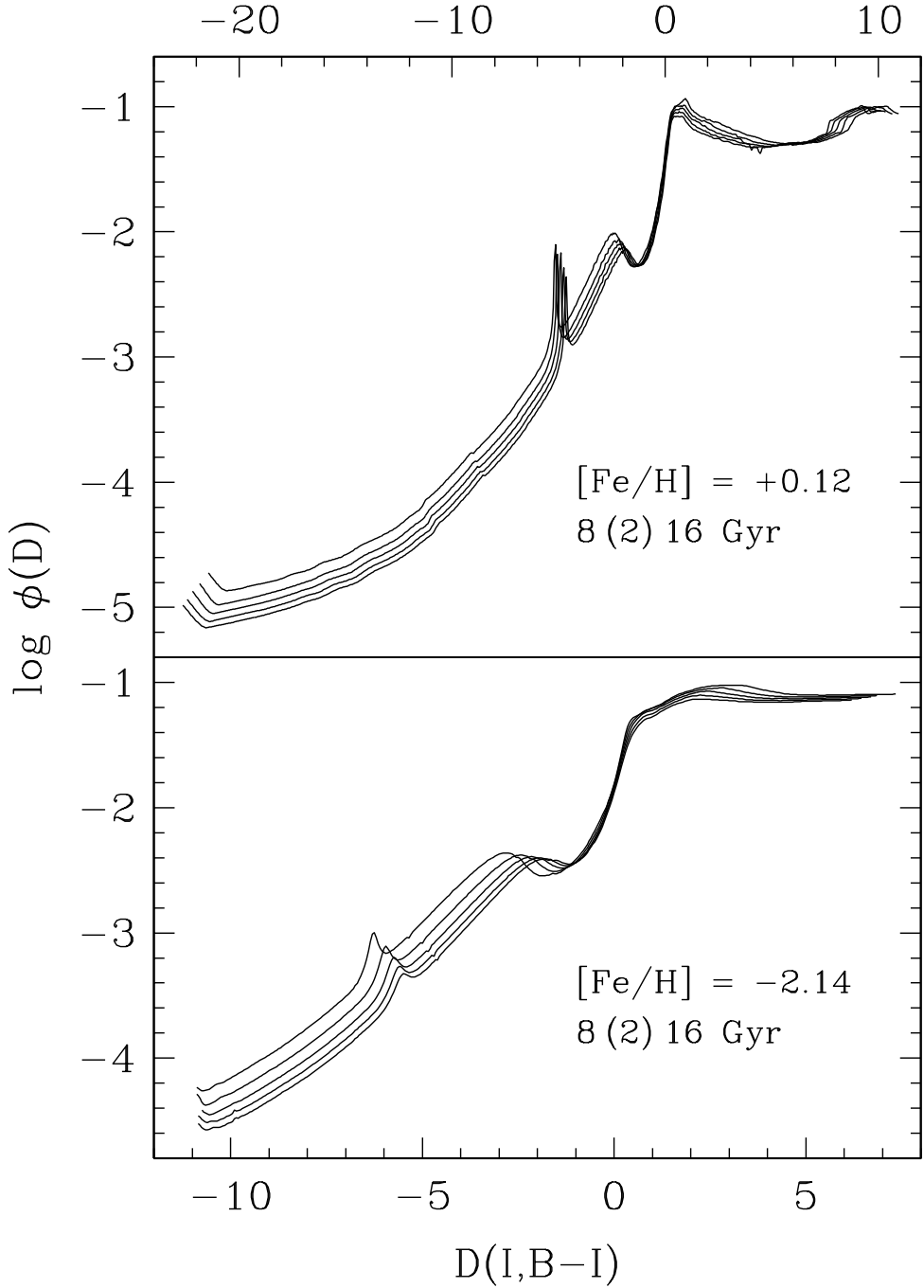}{7.5in}{0}{80}{80}{-252}{-54}
\caption{Differential IPFs in 0.05~mag distance bins are
plotted as a function of distance on the $I$--(\bi) observer's plane.
The very small wiggles on the lower giant branch portions of these IPFs
can be traced back to the evolutionary tracks; the larger, box-like
structures seen on the main sequence and upper giant branch of the
[Fe/H] $=+0.12$ models are artifacts produced by the color-temperature
transformations and by the bolometric corrections which have been applied.
\label{f5}}
\end{figure}

First we describe the general morphology of the IPFs with respect to
the expected evolutionary behavior, using the examples plotted in
Figure~\ref{f5}.  The transition from the main sequence to the RGB
occurs near $D(I,B\!-\!I)=0$ for both the metal-rich and the metal-poor
cases, because the distance, as we have defined it, is calculated with
reference to a point on the subgiant branch 0.05~mag redward of the
turnoff. The relatively broad local maximum immediately following the
transition, at $D(I,B\!-\!I)\approx -2$, corresponds to the inflection
point in the temperature evolution at the base of the RGB. The
evolutionary pause on the RGB is located at a distance near $-5$ to
$-6$ in both cases. The extension of the upper RGB to distances beyond
$-20$ in the $\rm [Fe/H]=+0.12$ grid occurs because giant branch
evolution in color is stronger than in more metal-poor stars, and color
evolution is weighted more strongly than luminosity evolution in the
calculation of distance along an isochrone.

In the theorist's plane, the derivatives of luminosity and temperature
with respect to time along evolutionary sequences tend to have some
numerical noise on the lower RGB, more obviously in the metal-rich grids.
The very fine wiggles seen on the lower RGB portion of the interpolated
metal-rich IPF plotted in the upper panel of Fig.~5 are a manifestation
of this.  

All of the other structures not predicted by canonical theory arise
from the adopted color--$\teff$ relations and the bolometric
corrections.  This is particularly evident at the faint end of the
models in the upper panel of Fig.~5, where the plateau-like structure
near $D=+10$ in the IPF has no counterpart at the faint end of the
theorist's IPFs plotted in Fig.~4.  Similar irregularities can be seen
on the upper RGB, such as the one that is weakly evident near $D=-10$
in the upper panel. It turns out that this anomaly is also found near
\bi$=3.4$ in the corresponding CF (not shown), and that it occurs over
the same range in (\bi) as the plateau-like structure on the main
sequence.  Moreover, artifacts of the bolometric corrections, evident
near the bright and faint ends of the corresponding LF (not shown), are
also found to occur over the same range in (\bi).

One conclusion that can be drawn from Fig.~5 is that neither the
color--$\teff$ relations nor the bolometric corrections are solely
responsible for a number of the features on the computed IPFs.  Most of
them are probably not real; indeed, it is our impression that the
transformations to the observed plane are not as consistently
determined for the metal-rich grid as for the metal-poor grid.  For
instance, only very minor artifacts appear on the IPFs for the latter
case. The small glitch near $D=-5$ on the IPF can be traced to a
corresponding glitch near \bi$=1.6$ on the CFs, etc. Comparisons with
observations are needed for definitive conclusions: at this point, we
can only say that, for the metal-rich grids in particular, the
discontinuous behavior on both the main-sequence and on the RGB
portions of the CFs between $3.2\lta B\!-\!I\lta 3.8$, and the
corresponding wiggles at the bright and faint ends of the LFs originate
in the same $\log \teff$ region of the transformation tables in which
the color indices and bolometric corrections are
interpolated.\footnote{We suspect that some improvement in the methods
used to interpolate in these tables could reduce the noise arising from
this source.  It is our intention to check into this possibility when
the extension of the present grids to [Fe/H] $>-0.3$ is prepared for
publication.}

\section{Tabulations of the Models}

For this paper, the grids of evolutionary tracks described in Paper I
have already been processed with software that automatically locates
the primary EEPs.\footnote{The software (written in FORTRAN 77) to produce
isochrones, IPFs, LFs, and CFs, as well as the input .eep files for the
evolutionary tracks presented in this series of papers are available
on request: contact P.~Bergbusch via email at bergbush@phys.uregina.ca.}
In this section, we describe how to use the software and we give examples
of the file formats currently available. 

Each grid of models is contained in an EEP file (denoted by the file
extension .eep) with the format illustrated in Table~\ref{tab1}. The
first seven lines have the same format in all of the subsequent files
produced from the EEP file. The first line specifies that this
particular file contains seven tracks; lines 2 through 6 list the
adopted chemical abundances and
line 7 gives the assumed mixing length parameter. For each track,
there is a header line listing the mass, the number of models in the
track ({\tt Npts}), the last model generated via the Lagrangian code
({\tt Match}), the shifts in age and in $\log T_{\rm eff}$ applied to
the subsequent giant branch models computed via the \cite{ppe71}
method\footnote{As described by \cite{van92}, the University of Victoria
code employs the usual Henyey method to solve the stellar structure
equations, with mass as the independent variable, when dealing with
evolutionary phases prior to the lower RGB.  Thereafter, the
non-Lagrangian technique developed by Eggleton (1971) is used (for
efficiency reasons) to extend the tracks to the tip of the giant branch.},
the age of the first model on the track, and the model numbers of the
primary EEPs. In the example illustrated, the primary EEPs for the $1.1
\cal M_\odot$ track are located at models 1, 36, 134, 198, 278, and 434
--- this track does not have a convective hook and thus the three spaces
following the turnoff point (36) are blank.  Within each track, the
entries are the model number, the luminosity, the effective
temperature, the age, the central hydrogen content, the derivative of
luminosity with respect to time, and the derivative of the effective
temperature with respect to time.

\begin{table}
\dummytable\label{tab1}
\end{table}

Isochrones on the $\log L$--$\log T_{\rm eff}$ plane are generated with
the program MKISO, which interactively prompts the user for the input
EEP file and a series of isochrone ages. Table~\ref{tab2} illustrates
the format of the isochrone files (denoted by the extension .iso)
generated.  The first line indicates that the file contains 6
isochrones. The header line for each isochrone lists its age and the
number of points generated.  The individual lines for each isochrone
list the luminosity, the effective temperature, the mass, the distance
along the isochrone, and two values of the derivative of the mass with
respect to distance along the isochrone. The first of these values is
computed via equation (2); the second is computed directly via five
point differentiation along the isochrone.

\begin{table}
\dummytable\label{tab2}
\end{table}

Isochrones, IPFs, LFs, and CFs are generated with the program MKIPF,
which prompts the user for the input isochrone file, the passband ($B$,
$V$, $R$, or $I$) for the magnitudes, the color index (\bv, \br, \bi,
\vr, or \vi), and three values for the exponent, $x$, in the power law
mass spectrum given by $N({\cal M})\propto {\cal M}^{(1+x)}$. The user
then specifies the type of output desired and is asked to specify the
bin size. For example, if the option to produce an IPF file (with the
extension .ipf) is chosen, the bin size is specified in terms of the
distance along the isochrone, in whatever passband and color index were
selected; if the LF option (with the extension .lfn) is chosen, the bin
size is specified in terms of the passband selected; if the CF option
(with the extension .cfn) is chosen, the bin size is specified in terms
of the color index selected.

The format of an IPF file is shown in Table~\ref{tab3}. In this
example, the magnitude passband and the color index selected were $I$
and \vi, respectively. The three values supplied for the mass-spectrum
exponent were $-0.5$, $+0.5$, and $+1.5$. For each age specified in the
input ioschrone file, the isochrones and IPFs are tabulated at equal
distance intervals. The entries on a line are the magnitude, the color
index, the mass, the bolometric magnitude, $\log\teff$, $\log g$, the
distance along the isochrone (the bin size is specified with respect to
this distance), and three pairs of differential and cumulative
(logarithmic) IPFs, one pair for each value of $x$ specified. The last
line for each age specified gives the parameters for a model at the RGB
tip.

\begin{table}
\dummytable\label{tab3}
\end{table}

The format of an LF file is shown in Table~\ref{tab4}. In this
example, both the differential and cumulative distributions are
tabulated at the centers of 0.2~mag bins, except for the bin at the RGB
tip. The format of a CF file is almost the same. However, since stars
in the post-turnoff stages of evolution have the same colors as those
along the main sequence, the distributions have been separated at the
color of the turnoff. For the example CF file shown in
Table~\ref{tab5}, there are 53 ({\tt Nms}) bins associated with the
main sequence in the 8 Gyr model; the post-turnoff portion begins at
bin 54.

\begin{table}
\dummytable\label{tab4}
\end{table}

\begin{table}
\dummytable\label{tab5}
\end{table}

\section{Additional Tests of the Models}

In order to make reliable stellar populations models, the predicted
luminosity, $\teff$, and color scales of the underlying evolutionary
tracks and isochrones must be as accurate as possible.  In this section,
we describe some further tests of the theoretical calculations, beyond
those presented in Papers I and II, which show that they seem to satisfy
most observational constraints rather well. 

\subsection{The Subdwarf $T_{\rm eff}$ Scale}

    As discussed in Paper I, the value of the mixing-length parameter,
$\alpha_{\rm MLT}$, that has been assumed in these calculations, was
set by the requirement that a Standard Solar Model reproduce the
observed temperature of the Sun. (To be consistent with the models
presented in this series of papers, the reference solar model did not
take diffusive processes into account.) In support of the resultant
$T_{\rm eff}$ scale, it was shown that the location on the ($M_V,\,\log
T_{\rm eff}$)--plane of the well-observed subdwarf HD$\,$103095
(perhaps better known by the name Groombridge 1830) is matched quite
well by models for the appropriate chemical abundances if its
temperature is $\gta 5100$ K.  Although there is considerable support
for this estimate (e.g., \cite{ki93}; \cite{gcc96}), many would argue
that something closer to 5000 K is more realistic (see, e.g.,
\cite{aam96}; \cite{bc96}).  Indeed, the effective temperatures of most
stars, including the best-studied Population II subdwarfs, appear to be
uncertain at the level of about 100 K, which clearly limits the ability
of such data to provide stringent constraints on predicted $T_{\rm
eff}$ scales.

    However, it is worthwhile to explore this issue further --- for the
following reason.  Paper I concluded that the present models provide a good
fit to the local subdwarfs on the [$M_V,\,(B-V)_0$]--diagram, {\it if}
metallicities close to those adopted by Gratton et al.~(1997) are assumed for
them.  But whether or not comparable agreement is found on the [$M_V,\,\log
T_{\rm eff}$]--plane, if the {\it temperatures} derived by Gratton et al.~are
assumed, was not investigated.  In fact, it turns out that there is quite good
consistency between these two plots.  To illustrate this, we have constructed
``mono-metallicity'' subdwarf sequences consisting of those subdwarfs that have
the smallest uncertainties in their parallaxes, along with a few subgiants to
permit an evaluation of the ages of stars that have evolved past the 
main-sequence turnoff.   That is, we have used our isochrones (only in a
differential sense) to correct the inferred $T_{\rm eff}$ and intrinsic color
of each field halo star, at its observed $M_V$, to the temperature and color
it would have if it had [Fe/H] = $-2.31$.  (We chose the lowest metallicity in
our grid as the reference [Fe/H] value because the few subgiants that we have
considered are very metal-deficient and it is desirable that the 
$T_{\rm eff}$/color adjustments which are applied to them be as small as
possible to avoid introducing significant errors into the determination of 
their ages.)

\begin{figure}
\figurenum{6}
\plotfiddle{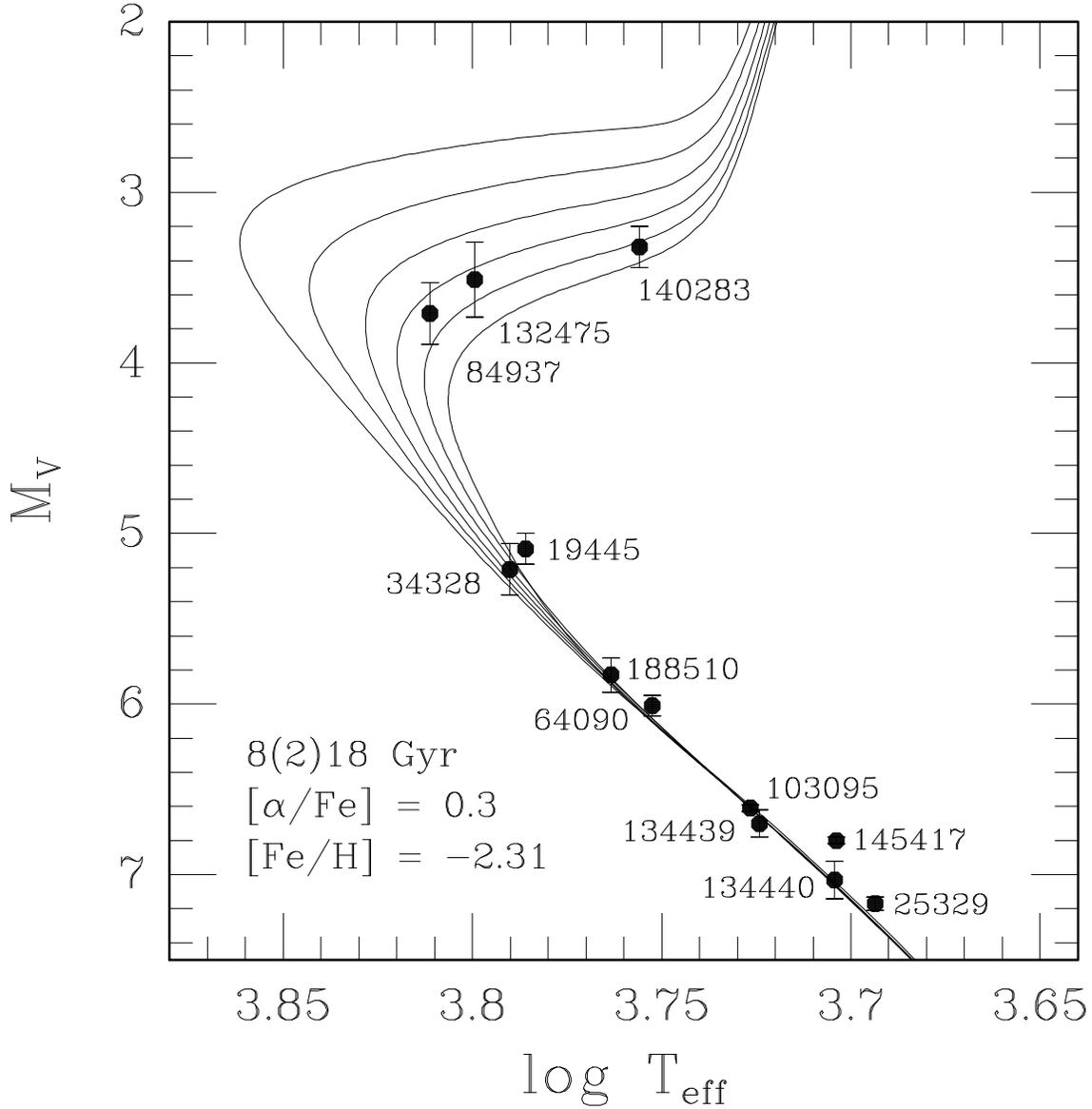}{6.25in}{0}{100}{100}{-306}{-108}
\caption{Comparison of isochrones for the indicated
parameters with the properties of several Population II subdwarfs and
subgiants (identified by their HD numbers), after the effective
temperatures of the latter were corrected to the values they would have
if their iron abundances corresponded to [Fe/H] $=-2.31$. For all but
one of the stars, the temperature adjustments that were applied to
produce the ``mono-metallicity'' subdwarf sequence were $\le 0.017$ in
$\log T_{\rm eff}$. Because the point representing HD$\,$132475
involved a rather large temperature adjustment ($\delta\log\teff =
0.036$), its location in this diagram is arguably the least well
determined of all of the stars that have been considered. The sources
of the stellar data are given in the text.  Note that Lutz-Kelker
corrections were applied to the $M_V$ values using the formula adopted
by Carretta et al.~(2000): these amounted to $\le 0.04$ mag, except in
the cases of HD$\,$84937 and HD$\,$132475, for which $\delta M_V^{\rm
LK} = -0.07$ and $-0.11$ mag, respectively. Neglecting these
corrections, which may not be appropriate for this highly-selected
sample of stars, would imply slight increases in the ages of the 3
subgiants.  Note, as well, that all of the latter appear to be very old
stars, with ages near 15--16 Gyr.\label{f6}}
\end{figure}

     The result of this exercise on the [$M_V,\,\log T_{\rm eff}$]--plane is
shown in Figure~\ref{f6}.  The $M_V$, $\sigma(M_V)$, and [Fe/H] values
that have been adopted are from the updated list of parameters given by
\cite{cgcf00},\footnote{Their entry for the parallax of the essentially
unreddened star HD$\,$188510 appears to be incorrect: the {\it Hipparcos}
catalogue gives $\pi = 25.32\pm 1.17$ mas for this star (instead of 22.80 mas),
which implies that its $M_V$ is $5.83\pm 0.10$, if $V=8.83$ is taken to be
its apparent magnitude and a Lutz-Kelker correction of $-0.02$ mag is applied.}
while the sources of the temperature data are Gratton et al.~(1996) and, in the
case of HD$\,$134439 and HD$\,$134440, \cite{cgcs99}.\footnote{Effective
temperatures for HD$\,$132475 and HD$\,$145417 were kindly provided to us by
Raffaele Gratton: they are, in turn, 5800 K and 4953 K.}  As the
$\alpha$-elements appear to be overabundant by $\sim 0.3$ dex in the vast
majority of field halo stars (see, e.g., the Clementini et al.~study), the
offsets in $T_{\rm eff}$ [and in $(B-V)_0$ color, to be discussed shortly] that
were applied to the selected stars were derived from our isochrones for
[$\alpha$/Fe] $=0.3$, an age of 14 Gyr, and the range in [Fe/H] between $-2.31$
and $-1.14$.  [With the exception of HD$\,$132475, the locations of stars in
Fig.~6 are nearly, or completely, independent of the age of the isochrones
which are used to determine the offsets.  Moreover, the position of HD$\,$132475
remains indicative of high age ($> 14$ Gyr), irrespective of which isochrones
are employed in this analysis.]

     The main conclusion to be reached from Fig.~6 is that the theoretical and
``observed'' $T_{\rm eff}$ scales at $M_V > 5$ agree extremely well.  This is
very encouraging from our perspective because, as discussed in their paper, the
Gratton et al.~(1996) determinations of effective temperature are consistent
with those derived from a number of methods --- including the fitting of Balmer
line profiles (e.g., \cite{fag94}), which is widely considered to be one of the
more accurate techniques.  It is interesting to note that HD$\,$19445 is
believed to have quite high oxygen and $\alpha$-element abundances ([O/Fe]
$=0.56$ and [$\alpha$/Fe] $=0.38$, according to Gratton et al.~1997), which may
explain why this star sits to the right of the isochrones.  On the other hand,
it is puzzling that the locations of HD$\,$134439 and HD$\,$134440 are not
discrepant, given that they are examples of the small minority of Population II
field stars that apparently have [$\alpha$/Fe] $\approx 0$ (\cite{ki97}).
However, we would not expect their displacements to hotter temperatures to be
more than $\delta(\log T_{\rm eff}) = 0.005$--0.010, judging from our models,
and such small shifts are probably consistent with the uncertainties associated
with the $T_{\rm eff}$ and [Fe/H] determinations.  (We have no explanation for
the somewhat anomalous location of HD$\,$145417 relative to the locus defined
by the other subdwarfs.)

\begin{figure}
\figurenum{7}
\plotfiddle{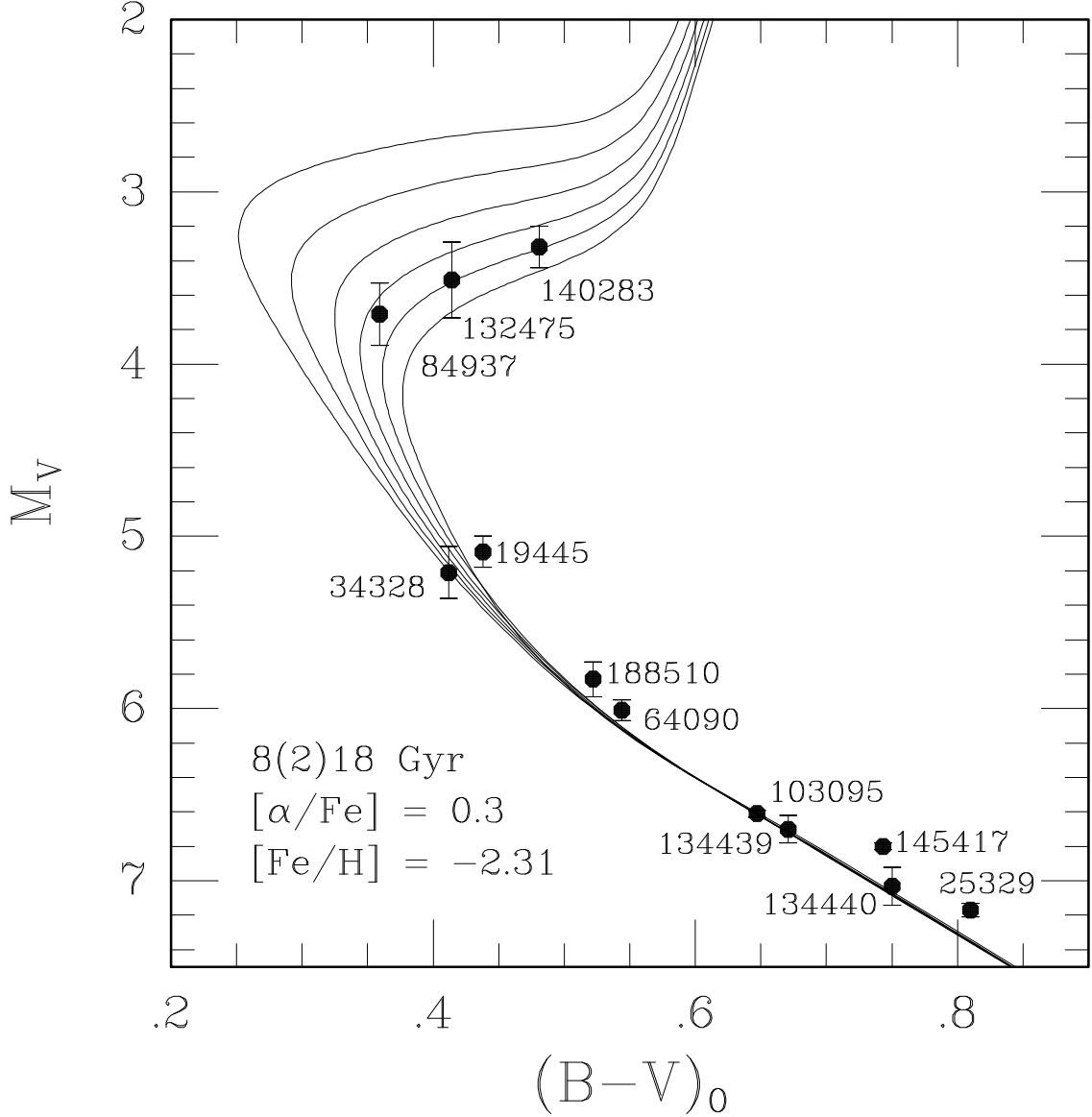}{6.5in}{0}{100}{100}{-306}{-108}
\caption{Similar to the previous figure, except that comparison
is carried out on the [$M_V,\,(B-V)_0$]--plane.  The observed colors and adopted
reddenings of the stars that have been considered are as given by Carretta et
al.~(2000).  In constructing the ``mono-metallicity'' subdwarf sequence, the
corrections that were applied to the intrinsic colors ranged from $+0.018$
mag (in the case of HD$\,$140283) to $-0.108$ mag (for HD$\,$132475, whose
position in this diagram is especially uncertain).\label{f7}}
\end{figure}

     As illustrated in Figure~\ref{f7}, the same subdwarfs define a
mono-metallicity sequence on the $[M_V,\,(B-V)_0]$--plane that is also in good
agreement with the same isochrones, when the predicted effective temperatures
are converted to $B-V$ colors using very close to \cite{bg89} transformations
(see Paper I for some discussion of the adopted color--$T_{\rm eff}$ relations).
Thus, there is no obvious need to adjust either the temperatures or the colors
of the models in order to satisfy the constraints posed by the subdwarf sample
that has been considered.  Granted, a couple of the stars (HD$\,$25329,
HD$\,$188510) are slightly displaced from the theoretical loci in Fig.~7, but
not in Fig.~6, which suggests that the adopted temperatures and colors are not
perfectly consistent with one another.  But this could simply be an indication
that the reddenings of these two stars are slightly higher than the values
adopted by Carretta et al.~(2000).   An underestimation of the reddenings of
HD$\,$134439 and HD$\,$134440 could also explain why they sit on the same locus
as HD$\,$103095 despite the latter having [$\alpha$/Fe] $\approx 0.3$ (Gratton
et al.~1997), and the former two stars having [$\alpha$/Fe] $\approx 0.0$, as
already noted.

     The other striking feature of Figs.~6 and 7 is that high ages are
implied by the locations of the post-turnoff stars.  As emphasized by
\cite{gru00}, HD$\,$140283 provides an especially strong argument that there
are very old stars and, furthermore, that the most metal-deficient globular
clusters must be of comparable age if the local subgiants are representative of
cluster subgiants of similar metal abundance.  Indeed, Paper II has shown that
excellent agreement between synthetic and observed C-M diagrams is obtained for
such systems as M$\,$92 if its age is $\sim 16$ Gyr.  To be sure, there is
sufficient flexibility in the Mixing Length Theory of convection to obtain good
fits to the morphology of an observed C-M diagram for any assumed distance
(within reason).  But if M$\,$92 is as young as 12 Gyr, say, then the zero-point
of our adopted color--$T_{\rm eff}$ relations would require a large correction,
in conflict with our findings from Figs.~6 and 7, and M$\,$92 would be $\sim
4$ Gyr younger that field halo stars of similar [Fe/H], which seems unlikely.

\begin{figure}
\figurenum{8}
\plotfiddle{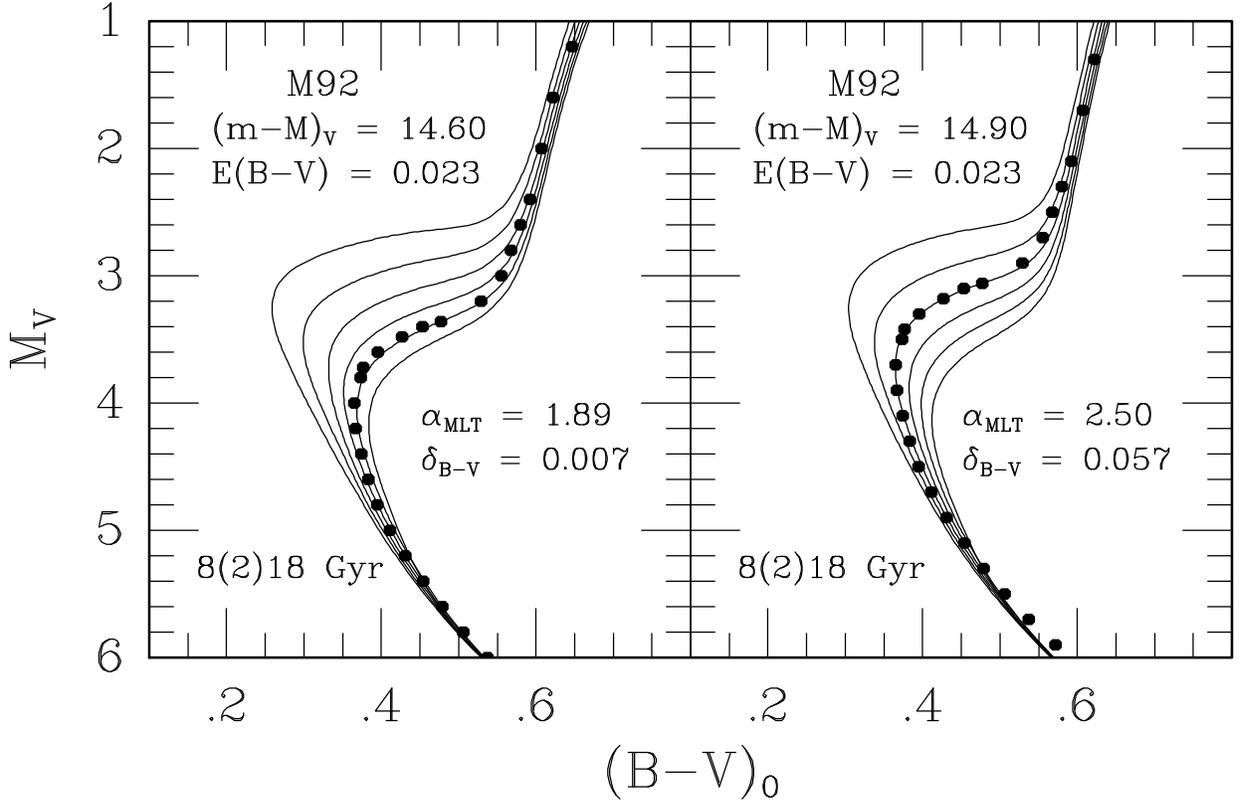}{3.5in}{0}{100}{100}{-306}{-144}
\caption{Fits of isochrones for [Fe/H] $=-2.31$ and [$\alpha$/Fe]
$=0.3$ to the Stetson \& Harris (1988) fiducial for M$\,$92, supplemented by
data for the lower giant branch from M.~Bolte (private communication).
The Schlegel et al.~(1998) reddening estimate has been assumed.  The
only difference in the models is the value that has been assumed for the usual
mixing-length parameter, $\alpha_{\rm MLT}$.  The main point of this figure is
that it is possible to obtain comparably good agreement between synthetic and
observed C-M diagrams on the assumption of distance moduli that differ by 0.3
mag (and ages that differ by $\approx 4$ Gyr) simply by choosing the value of
$\alpha_{\rm MLT}$ appropriately.  However, very different zero-point offsets
must be applied to the colors of the isochrones in order to match the observed
turnoffs in the two cases and, on this basis, the comparison given in the 
left-hand panel is clearly favored over that given in the right-hand panel.\label{f8}}
\end{figure}

     To elaborate on this point, we show in Figure~\ref{f8} two fits of
isochrones to the M$\,$92 C-M diagram where the only difference in the models
is the choice of $\alpha_{\rm MLT}$, the usual mixing-length parameter.  The
left-hand panel reproduces the same isochrones that were plotted in Fig.~7 and
it is obvious that a superb match to the cluster fiducial is obtained if the
adopted distance is near $(m-M)_V = 14.60$ and if $E(B-V) = 0.023$ mag
(\cite{sfd98}).  This case assumes a value of
$\alpha_{\rm MLT}$ that is needed to satisfy the solar constraint (see Paper I),
and the distance is approximately what one would infer from HD$\,$140283 if
M$\,$92 and the field subgiant are close to being coeval.  However, even if the
cluster distance modulus is actually as high as $(m-M)_V = 14.90$, it is still
possible to obtain an equally fine fit to the {\it shape} of the observed C-M
diagram simply by adopting $\alpha_{\rm MLT} = 2.50$ (see the right-hand
panel).\footnote{Discrepancies do become evident at both brighter and fainter 
magnitudes than those plotted if the high value of $\alpha_{\rm MLT}$ is 
assumed.  Hence, in this instance, one may have to postulate some variation in
the value of the mixing-length parameter with evolutionary state in order to
match the entire C-M diagram.  Of course, there is no reason why this parameter
should be completely independent of $T_{\rm eff}$ and/or $\log g$, but it is
interesting that no such difficulty is found if $\alpha_{\rm MLT} = 1.89$.}
The main problem with this scenario is that a very large correction must be
applied to the model colors in order for the relevant isochrone (for an age of
$\approx 12$ Gyr) to match the observed turnoff color.  Such a large color 
shift would seem to be precluded by the comparisons given in Figs.~6 and 7.

\subsection{On the Globular Cluster and Subdwarf [Fe/H] Scales}

     While Paper I and the present discussion appear to offer strong support
for the $T_{\rm eff}$ and [Fe/H] scales derived by Gratton et al.~(1996) and
Carretta et al.~(2000) for the local subdwarfs, Paper II has suggested that
the metallicities determined by \cite{cg97} for globular clusters, especially
those with $-2 \lta$ [Fe/H] $\lta -1$, are too high.  Something much closer to
the \cite{zw84} scale was favored.  Because this is such an unexpected (and
controversial) result, some additional comments on this matter are worthwhile.
In particular, if our inference from Figs.~6 and 7 is correct, that our isochrones
require little or no adjustment of the predicted colors to match the local
calibrators, then we should obtain reliable estimates of globular cluster
distances by performing main-sequence fits of the cluster fiducials to our
isochrones.  Quite understandingly, the results of such an exercise depend
sensitively on what metal abundances are assumed.

\begin{figure}
\figurenum{9}
\plotfiddle{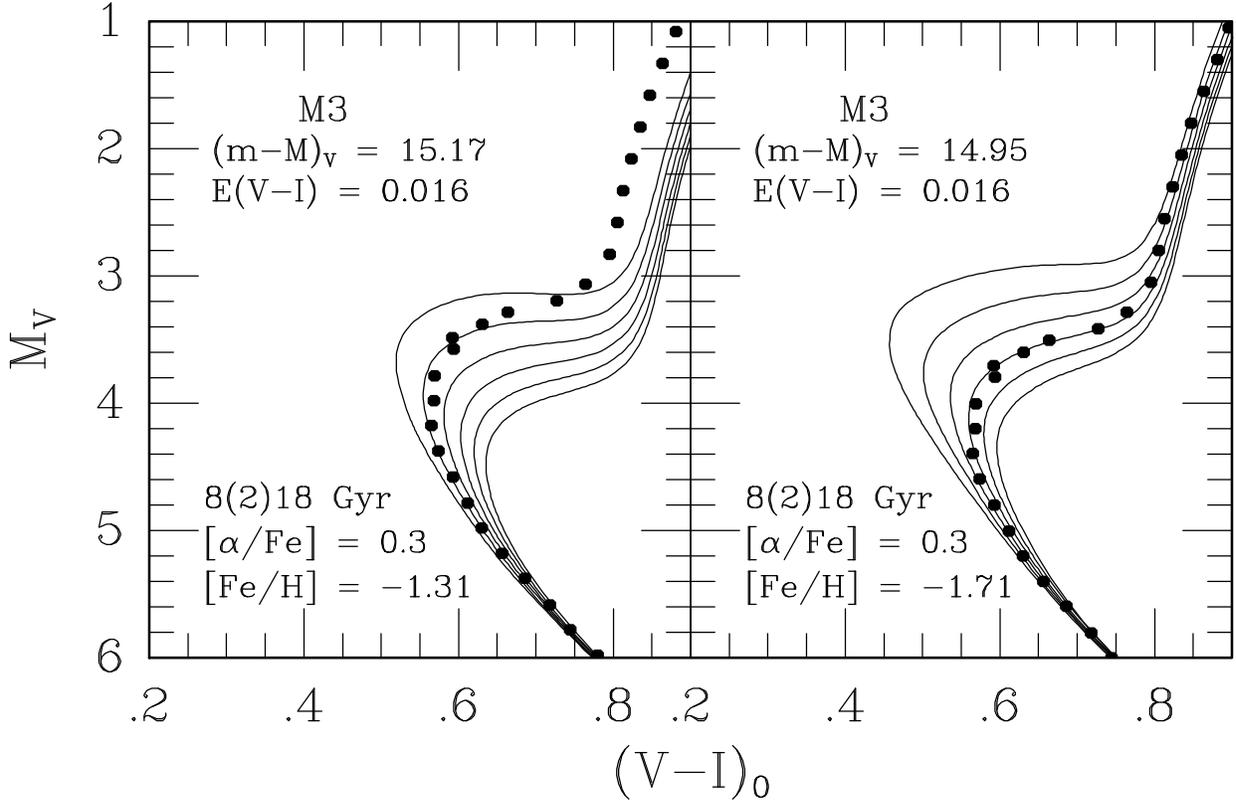}{3.5in}{0}{100}{100}{-306}{-144}
\caption{Main-sequence fits of the Stetson et al.~(1999) fiducial
for M$\,$3 to isochrones for the indicated parameters.  The adopted reddening
is consistent with that given by Schlegel et al.~(1998).\label{f9}}
\end{figure}

     This is illustrated in Figure~\ref{f9}, using M$\,$3 as an instructive
example.  Recall that Carretta \& Gratton (1997) have determined [Fe/H] $=-1.34$
for this system whereas the Zinn \& West (1984) estimate is $-1.66$.  Hence, the
most appropriate isochrones in our grid to use in this analysis are those having
[Fe/H] $=-1.31$ and $-1.71$.  If the \cite{stet99} fiducial for M$\,$3 is 
fitted to the models of higher metallicity (see the left-hand panel), then one
obtains $(m-M)_V = 15.17$ and an age of $\approx 10$ Gyr.  (A main-sequence fit
to local subdwarfs will necessarily yield the same relatively young age for
M$\,$3 if the cluster iron abundance is as high as the Carretta \& Gratton
estimate.)  This is to be compared with $(m-M)_V = 14.95$ and an age of $\approx
14$ Gyr if a fit is  performed to the lower-metallicity isochrones (see the
right-hand panel). [Our transformations to $V-I$ are precisely as given by Bell
\& Gustafsson (1989) and, as shown in Papers I and II, very similar
interpretations of both cluster and subdwarf C-M diagrams are generally found
regardless of whether the color used is $B-V$ or $V-I$.]

     The isochrones for [Fe/H] $= -1.31$ clearly fail to match the morphology
of the M$\,$3 fiducial whereas the predicted and observed loci agree extremely
well if the assumed [Fe/H] is $-1.71$.  By itself, this is not a particularly
strong argument against the comparison shown in the left-hand panel of Fig.~9,
given that isochrone shapes are so easily manipulated
(as we have already demonstrated).  A more compelling
reason for rejecting this possibility is the fairly strong case that has been
made in support of an age of $\sim 16$ Gyr for M$\,$92 and the finding that
isochrones for this age, and a metallicity within the uncertainties of most
determinations, do not require any {\it ad hoc} adjustments to either the model
temperatures or colors to achieve nearly perfect coincidence with the cluster
(and subdwarf) C-M diagrams.  It seems highly improbable that M$\,$3 is 5--6 Gyr
younger than M$\,$92 and that the isochrone fit would would be so problematic
for the one cluster, but not for the other.  There are no such difficulties if
the [Fe/H] value obtained by Zinn \& West (1984) for M$\,$3 is more realistic
than that derived by Carretta \& Gratton (1997).

     This conclusion does depend on the accuracy of the assumed
     reddenings and $\alpha$-element abundances as well as, among other
things, the photometric zero-points, so it may well be revised somewhat
as further progress is made.  However, it would not be too surprising
if globular cluster abundances, as derived from spectroscopic data for
the brightest giants, are not quite on the same scale as that of local
field halo dwarfs.  (Note that the age estimates discussed above should
probably be reduced by about 10\% to take into account the effects of
diffusion, though current diffusive models are not without their
difficulties: for some discussion of this point, see Grundahl et
al.~2000.\footnote{After this paper was submitted for publication, a
preprint became available which adds to the evidence that something
inhibits the gravitational settling of heavy elements in at least the
surface layers of very metal-deficient stars.  \cite{gbb01} have found,
from VLT high-dispersion spectroscopy, essentially no difference in the
derived [Fe/H] values for large samples of turnoff and lower RGB stars
in two globular clusters, NGC$\,$6397 and NGC$\,$6752.  The failure to
find any variation of [Fe/H] across the cluster subgiant branches,
together with the unexpectedly high Li abundances found in both cluster
(\cite{mp94}) and field turnoff stars (e.g., \cite{ss82}, \cite{sw95}),
provide extremely important constraints on diffusion in Population II
stars that have yet to be properly evaluated.  It seems very likely
that the age effect, which arises from the settling of helium in the
deep interior, will persist, but the effective temperatures predicted
by currently available diffusive models may well need to be revised
significantly.  A collaboration between the University of Montreal
(G.~Michaud and colleagues) and D.A.V. is underway to study this
problem.})

\subsection{RGB Slopes and RGB Tip Magnitudes}

\begin{figure}
\figurenum{10}
\plotfiddle{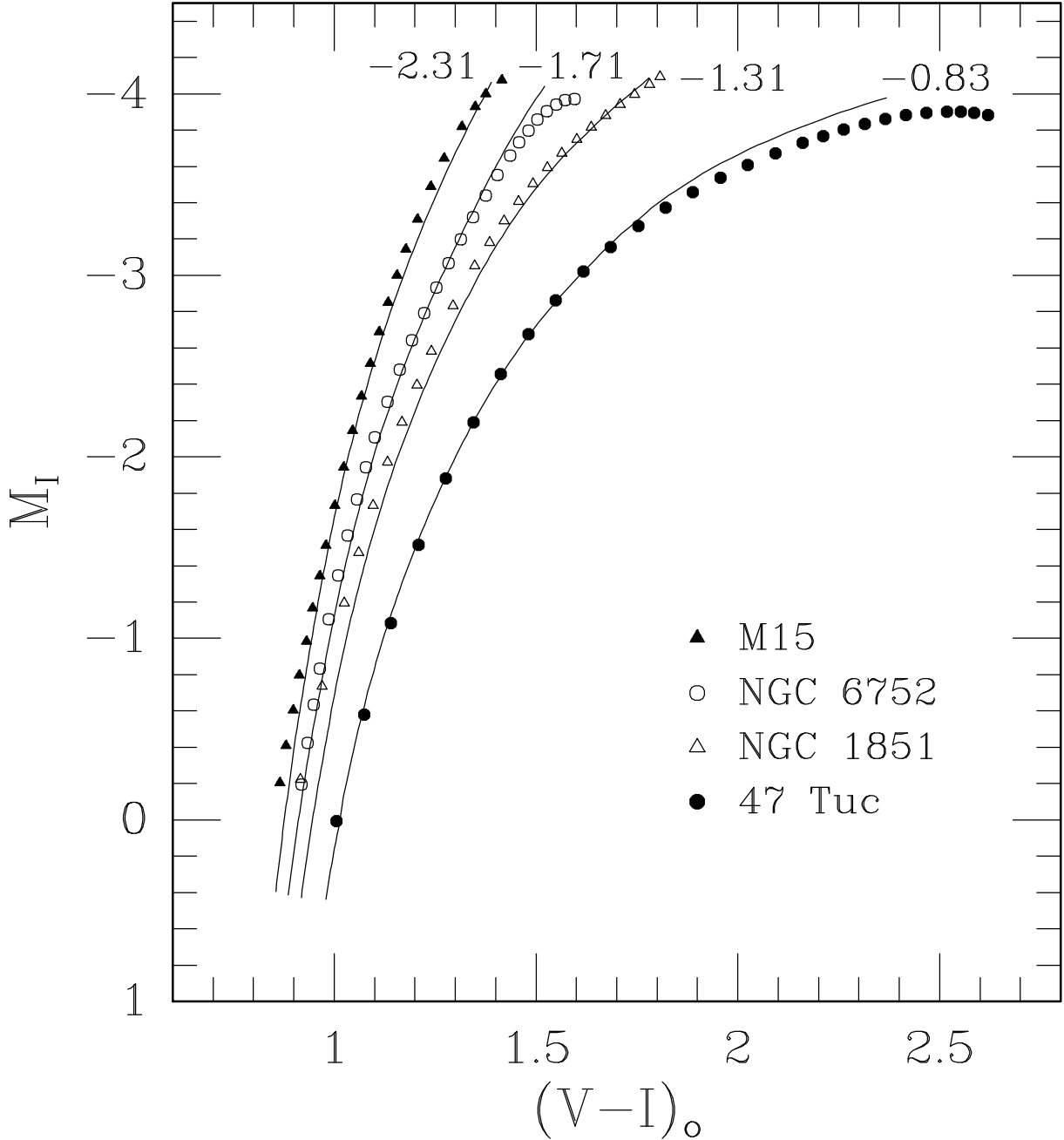}{7.25in}{0}{100}{100}{-306}{-100}
\caption{Overlay of the giant-branch segments of isochrones for
the indicated [Fe/H] values onto the fiducial sequences derived by Da Costa \&
Armandroff (1990) for M$\,$15, NGC$\,$6752, NGC$\,$1851, and 47 Tuc.  For these
four clusters, in turn, we have assumed $E(B-V) = 0.108, 0.056, 0.034,$ and
0.032 mag (Schlegel et al.~1998), as well as  $(m-M)_V = 15.43, 13.25, 15.55,$
and 13.37 mag (see Paper II), which imply cluster ages near 14, 13, 11.5, and
11.5 Gyr, respectively (see Paper II).  To produce this plot, $A_V =
3.1\,E(B-V)$, $A_I = 0.59\,A_V$, and $E(V-I)= 1.25\,E(B-V)$ have also been
assumed (see, e.g., Bessell et al.~1998).\label{f10}}
\end{figure}

     Paper I has already shown that, both in zero-point and slope, the computed
giant branches on the ($M_{\rm bol},\,\log T_{\rm eff}$)--plane agree very
well with the loci inferred for globular clusters from infrared photometry.  Our
intention here is merely to point out that reasonably good agreement between
theory and observations is also found on the various C-M diagrams that can
be constructed from $BV(RI)_C$ magnitudes.  As an example, we show in
Figure~\ref{f10} how well our isochrones are able to reproduce the
\cite{dca90} fiducials for M$\,$15, NGC$\,$6752, NGC$\,$1851, and 47 Tuc on the
[$M_I,\,(V-I)_0$]--diagram.  In preparing this figure, we have adopted the
reddenings derived by Schlegel et al.~(1998), relations giving $E(V-I)$ and
$A_I$ as a function of $E(B-V)$ from \cite{bcp98}, and the distance moduli that
were derived in Paper II; thus, the fiducials tabulated by Da Costa \&
Armandroff have been have corrected to take the small differences in $E(B-V)$
and $(m-M)_V$ between their adopted parameters and ours into account.  Although
we have plotted the giant-branch segments of isochrones for certain ages,
to be consistent with the assumed distances, the comparison is fairly
independent of this choice.  It should be noted, as well, that the implied
metallicities of the four clusters are within 0.1 dex of the [Fe/H] values
listed in the compilation of such data by \cite{har96}.

     At [Fe/H] $\lta -1.5$, the adopted transformations to $V-I$ are exactly
as given by Bell \& Gustafsson (1989), whereas some adjustment of their
color--$T_{\rm eff}$ relations (only at low gravities) was required at higher
metal abundances to achieve consistency between the theoretical and observed
planes.  These corrections were always to the red and they tended to be
systematically larger at higher [Fe/H] values, which may indicate the lack of
sufficient blanketing in the model atmospheres constructed by Bell \& Gustafsson
for cool, relatively metal-rich giants.  In any case, it is desirable to force
the {\it colors} of the upper giant branches to agree with those observed in
order to enhance the usefulness of our models in stellar population studies.
[Even though there are some discrepancies between the [Fe/H] $=-0.83$ isochrone
and the 47 Tuc fiducial at the brightest magnitudes, this does not necessarily
indicate a problem with the adopted color transformations.  It may be, for
instance, that 47 Tuc has a slightly higher iron abundance, and given the huge
sensitivity of the $V-I$ color index (as well as the bolometric corrections)
to $T_{\rm eff}$ at low temperatures, a better fit to the upper end of the
cluster fiducial would be obtained if either the predicted temperatures were
reduced slightly or the adopted [Fe/H] value were increased by $\lta 0.1$ dex.]

     In contrast to the colors, which were constrained to some extent to satisfy
empirical data (for giants only), the bolometric corrections to $V$ magnitudes
do represent purely theoretical predictions (at all temperatures, gravities, and
[Fe/H] values).  In particular, we have adopted the bolometric corrections
given by \cite{wb94}, which are based on Kurucz model atmospheres and which 
span a very wide range in parameter space.  They agree extremely well in a
systematic sense with those computed recently by R.A.~Bell (see, e.g., the
discussion by \cite{vi97}).  For instance, we have found that Bell's
transformation of selected isochrones to fit the M$\,$92 and NGC$\,$2419 C-M
diagrams in the study by \cite{har97} is nearly indistinguishable with ours.
Hence, a comparison of predicted and observed giant-branch tip magnitudes does
provide a valid test of the theoretical models.  (Bolometric corrections in $I$
were obtained from $BC_I = BC_V + V-I$.)

     Considering that cluster distance moduli are currently uncertain by at
least 0.2 mag, the luminosity at which giant-branch evolution is predicted to
terminate, and its variation with metallicity, is in satisfactory agreement
with observations (see Fig.~10).  If anything, slightly shorter distances,
implying somewhat higher ages, may be indicated for M$\,$15 and NGC$\,$1851.
It should be appreciated that the {\it bolometric} magnitudes for the
giant-branch tip are within 0.05 mag of those predicted by A.V.~Sweigart for
stars of the same mass and chemical composition (see Paper I) and within
$\approx 0.1$ mag of those reported by e.g., \cite{ccdw98}, when very similar
physics is adopted.  (Sweigart's models are fainter than ours while those by
Cassisi et al.~are brighter.)  As the same neutrino cooling rates have been
used by the aforementioned investigators, we suspect that small differences in
the respective equation-of-state formulations are responsible for the minor
variations in the predicted tip magnitudes.  This should be investigated
further.\footnote{We note that
\cite{fmof00} have recently presented extensive, high-quality $JK$
photometry for the red-giant and horizontal-branch populations in 10 globular
clusters.  We hope to examine how well our models are able to reproduce their
observations at some future date, once we have fully investigated the
transformations to $V-K$ and $J-K$.}
 
\subsection{RGB Bump Magnitudes}

     Another important diagnostic in observed C-M diagrams is the luminosity
on the red-giant branch where the H-burning shell in evolving stars passes
through the chemical abundance discontinuity left behind when the convective
envelope reached its maximum penetration (near the base of the RGB).  When this
occurs, the evolution up the giant branch either slows down or reverses 
direction for a short time as the structure of the star adjusts to a somewhat
larger fuel supply (higher H abundance).  This manifests itself as a local
enhancement in the number of stars in an otherwise smoothly decreasing
differential luminosity function for the giant-branch component of a populous
system such as a globular cluster --- where, with few exceptions (e.g., $\omega$
Cen), the member stars are essentially coeval and chemically homogeneous.  As
discussed by \cite{cs97}, the luminosity of this so-called ``RGB bump'' is a
strong function of metallicity and age, and depends only weakly on other factors
such as the helium abundance and the mix of heavy elements.  However, one must
be wary of placing too much reliance on this feature as a distance (and age)
constraint, given the possibility that there may be some amount of overshooting
below the convective envelope (e.g., \cite{al91}), or that opacities in the
critical $\sim 10^6$ K regime are still inadequate, or that oxygen abundances
are much higher than we have assumed.  (As shown in Paper I, oxygen is the
dominant contributor to the opacity at $5.8 \lta \log T \lta 6.7$, and its
abundance will therefore have important consequences for the depth of the
envelope convection zones in giants; also see VandenBerg \& Bell 2001).

     Table~\ref{tab6} lists the bump magnitudes predicted by our models for
ages from 10 to 16 Gyr and the entire range of [Fe/H] and [$\alpha$/Fe] values
considered in this investigation.  The considerable sensitivity of
$M_V^{\rm bump}$ to these three parameters is clearly evident, and a careful
inspection of this table, in conjunction with Table 2 of Paper I (which gives
the $Z$ value for each mixture), reveals that the chemical composition
dependence is primarily with $Z$, the total mass-fraction abundance of the
elements heavier than helium.  That is, the detailed distribution of the metals
affects the results at only the level of a few to several hundredths of a
magnitude.  Compare, for instance, the tabulated magnitudes for [Fe/H] $=-2.31$
and [$\alpha$/Fe] $=0.6$ ($Z=3.07\times 10^{-4}$) with those for [Fe/H] $=-1.84$
and [$\alpha$/Fe] $=0.0$ ($Z=3.0\times 10^{-4}$): the differences in
$M_V^{\rm bump}$ are $\lta 0.04$ mag.   Only at relatively high values of $Z$
do these differences rise to $\approx 0.1$ mag; e.g., compare the tabulated
magnitudes for [Fe/H] $=-1.31$ and [$\alpha$/Fe] $=0.6$ ($Z = 3.07\times
10^{-3}$) with those for [Fe/H] $=-0.83$ and [$\alpha$/Fe] $=0.0$ ($Z =
3.0\times 10^{-3}$).

\begin{table}
\dummytable\label{tab6}
\end{table}

     At the same $Z$, the bump magnitudes predicted our models agree quite 
well (generally to within $0.05$ mag) with those reported by Cassisi \&
Salaris (1997, see their Table 1) and with those computed by \cite{scl97}, as
noted by \cite{rcp99}.  (Some differences can be expected due to the fact that
the different studies assume slightly different helium contents and 
heavy-element mixtures.)  Thus, the interpretation of this feature in observed
C-M diagrams should be nearly independent of whose models are used.  Much more
critical is the adopted metallicity.  For instance, Rood et al.~have found that
our models for [Fe/H] $=-1.31$, [$\alpha$/Fe] $=0.3$, and an age of $\approx
12$ Gyr accurately predict the location of the RGB bump in M$\,$3 (as do the
models by Salaris et al.~for the same age and nearly the same $Z$, but with the
metals in solar proportions).  In this cluster, $V_{\rm bump} = 15.45\pm 0.05$
(\cite{fer99}), and an age of 12 Gyr is obtained if the cluster distance modulus
is close to $(m-M)_V = 15.03$, which implies $M_V^{\rm bump} \approx 0.42$, in
excellent agreement with the relevant entry in Table 6.  However, as we have
argued above, such a high metal abundance for M$\,$3 is problematic.  Indeed,
Rood et al.~have found that it is impossible to choose a distance and
metallicity such that the corresponding models are able to reproduce the entire
luminosity function (LF).  In particular, they found that a good fit to the
subgiant LF resulted in a poor fit to the luminosity of the RGB bump, and vice
versa.  An appreciably higher age ($\gta 14$ Gyr), and/or a lower metallicity,
is indicated by the slope of the LF between the turnoff and lower giant branch.

    It is especially unfortunate that the chemical composition of M$\,$3 is
so uncertain at the present time: the [Fe/H] values determined by Zinn \& West
(1984), Carretta \& Gratton (1997), and \cite{kra98} differ from one another by
$\gta 0.15$ dex and span a range $>0.30$ dex.  Until these differences are
understood, it will be difficult to say whether or not models accurately predict
the observed bump magnitude in this cluster.  However, it does seem unlikely
that such a reconciliation of theory and observation is possible in the case
of M$\,$92, {\it if} it is older than $\sim 12$ Gyr.  For this system, current
metallicity determinations are much more consistent with one another --- e.g.,
high-resolution spectroscopy of bright giants by \cite{sne91} and Carretta \&
Gratton (1997) has yielded [Fe/H] $=-2.25$ and $-2.16$, respectively, while
Zinn \& West give [Fe/H] $=-2.24$.  Moreover, most spectroscopic studies seem
to find [$\alpha$/Fe] $\approx 0.3$ for most globular clusters, including
M$\,$92 (see the spectroscopic studies mentioned above, as well as the reviews
by Kraft 1994 and Carney 1996).\footnote{Should it turn out that M$\,$92 stars
have [O/Fe] $\approx 1.0$ (Israelian et al.~1998, Boesgaard et al.~1999),
then the cluster age could be as low as $\sim 13$ Gyr {\it and} there would
be no discrepancy between the predicted and observed bump magnitudes (see
VandenBerg \& Bell 2001).}

    As shown in Table~\ref{tab7}, our models are able to match the luminosity 
of the RGB bump in M$\,$92 only if the cluster metallicity is at the high end
of the observed range and its age is $\lta 12$ Gyr.  According to Ferraro et
al.~(1999), $V_{\rm bump} = 14.65\pm 0.05$ in M$\,$92, which corresponds to
the absolute visual magnitudes given in the fifth column if the distance moduli
in the third column are assumed.  At these distances, the ages obtained from
our models and the corresponding bump luminosities (from Table 6) are as
tabulated in the fourth and sixth columns, respectively.  The last column, which
contains the differences between the observed and predicted bump $M_V^{\rm bump}$
values, shows that the latter are too bright by 0.05 to 0.31 mag, depending on
the distance and iron abundance that is assumed.  In view of the fairly strong
case that has been made in support of an age near 16 Gyr for M$\,$92 (see
\S 4.1 above, and Grundahl et al.~2000), we are inclined to conclude that the
our models fail to predict the luminosity of the RGB bump in this cluster by at
least 0.2 mag.

\begin{table}
\dummytable\label{tab7}
\end{table}

     In fact, a very similar conclusion is reached from an analysis of the C-M
diagram for 47 Tuc, which has [Fe/H] $\gta -0.8$.  It has $V_{\rm bump} = 14.55
\pm 0.05$ mag (Ferraro et al.~1999), and if we adopt $(m-M)_V = 13.4$, which is
within 0.1 mag of most estimates (and slightly higher than our preferred value:
see Paper II), then $M_V^{\rm bump} = 1.15$.  At this distance, the age and
bump magnitude predicted by our isochrones for [Fe/H] $= -0.83$ and
[$\alpha$/Fe] $=0.3$ (\cite{bwo90}) are $\approx 11$ Gyr and 0.91 mag,
respectively.  The corresponding numbers for isochrones having [Fe/H] $=-0.71$
and the same $\alpha$/Fe ratio (Carretta \& Gratton 1997) are $\approx 10.5$ Gyr
and 1.00 mag.  Thus, the discrepancy between the predicted and observed values
of $M_V^{\rm bump}$ is, in turn, 0.24 and 0.15 mag.  Although the agreement
could be improved by adopting a larger distance, any age younger than $\approx
10$ Gyr at these metallicities poses problems for other features in the observed
C-M diagram.  For instance, the subgiant branch in 47 Tuc has a gentle upward
slope between the turnoff and lower RGB on the [$V,\,(B-V)$]--diagram, while
isochrones for ages $\lta 10$ Gyr predict that the variation in $V$ across the
subgiant branch is non-monotonic (see Fig.~34 in Paper II). 

\begin{figure}
\figurenum{11}
\plotfiddle{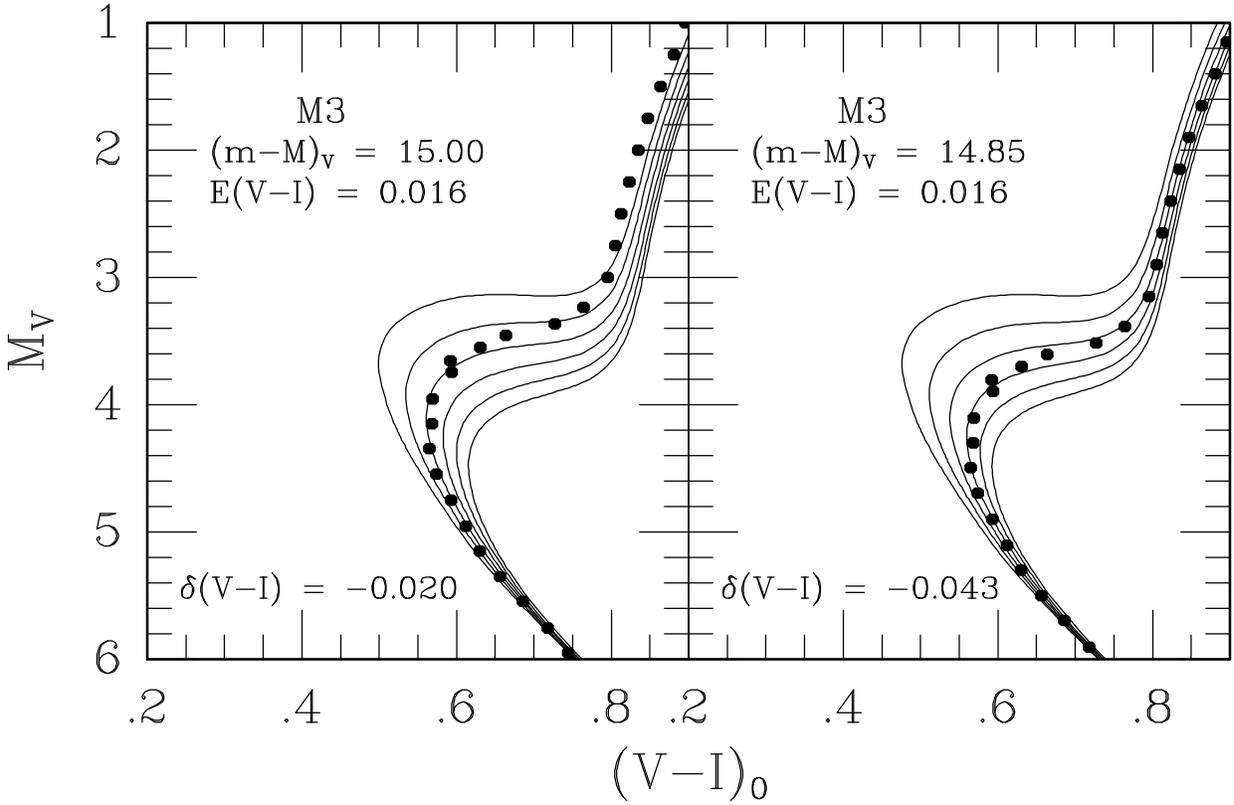}{3.5in}{0}{100}{100}{-306}{-144}
\caption{Similar to the left-hand panel of Fig.~9, except that
the cluster distance modulus has arbitrarily been set to the indicated values.
To reconcile the observed turnoff with that of the most appropriate isochrone,
the zero-point shift specified in the lower left-hand corner of each panel
was applied to the isochrone colors.\label{f11}}
\end{figure}

     If our models truly are incapable of explaining the location of the RGB
bump in both M$\,$92 and 47 Tuc, then it seems unlikely that they would fare any
better at intermediate metal abundances.  That is, there is some basis for
expecting that the bump luminosity in M$\,$3 {\it should} be $\sim 0.2$ mag
fainter than that predicted by the most appropriate isochrone in our grid.  It
turns out to be quite difficult to force such a discrepancy unless the cluster
is more metal poor than [Fe/H] $=-1.3$.  This becomes readily apparent when one
considers Figure~\ref{f11}, which illustrates how well isochrones for [Fe/H]
$=-1.31$ and [$\alpha$/Fe] $=0.3$ are able to reproduce the M$\,$3 C-M diagram,
on the assumption of smaller distances than that derived from a main-sequence
fit (see the left-hand panel of Fig.~9).  To identify which isochrone has the
same turnoff luminosity as the cluster, small offsets to the synthetic colors
were applied (as indicated).\footnote{The fact that we need to shift the
isochrones to the blue is itself suggestive that M$\,$3 has a lower [Fe/H],
because little or no adjustment to the model colors seems to be necessary to
match the properties of the local subdwarfs (as discussed above and in Paper I).
However, as noted in Paper II (see Fig.~12 therein), the three best photometric
data sets currently available for M$\,$3 have zero-points that differ by up to
$\approx 0.015$ mag in $V-I$: the Stetson et al.~(1999) fiducial, which we have
used, is the bluest, while those by \cite{jb98} and Rood et al.~(1999) are
redder.  Had we used either of the latter, the offsets applied to the isochrones
in Fig.~11 would have been smaller in an absolute sense by 0.01--0.015 mag (and
a smaller distance moduli by $\sim 0.08$ mag would have been derived in
Fig.~9).  The accuracy of photometric zero-points is obviously an important
concern in the main-sequence fitting technique.}

     Since $V_{\rm bump} = 15.45$, the assumption of $(m-M)_V = 15.00$ and
14.85 imply that $M_V^{\rm bump} = 0.45$ and 0.60 mag, respectively.  The
observed turnoff luminosities in these two cases are clearly very close to those
predicted by 12 and 14 Gyr isochrones, for which the predicted bump magnitudes
(see Table 6) are, in turn, 0.45 and 0.52 mag.  Because of the substantial 
dependence of $M_V^{\rm bump}$ on age, the difference between the theoretical
and observed bump luminosities varies relatively slowly with distance modulus.
Thus, in order to obtain $\Delta M_V^{\rm bump} \gta 0.15$ mag, M$\,$3 would
have to be older than 16 Gyr, if it has [Fe/H] $\approx -1.3$ and [$\alpha$/Fe]
$\approx 0.3$.  But such a high age would imply that the isochrone colors need
to be corrected blueward by $> 0.06$ mag in $V-I$, which is completely
unacceptable.  There is no such difficulty if a metallicity close to the Zinn
\& West (1984) estimate is assumed (see the right-hand panel of Fig.~9).  Note,
as well, how much better the isochrones for lower $Z$ match the observed C-M
diagram.  Interestingly, the 14 Gyr isochrone for [Fe/H] $=-1.31$ and
[$\alpha$/Fe] $=0.3$ {\it is} able to reproduce both the location of the main
sequence and the RGB, once a large zero-point adjustment is applied to the
synthetic colors, but not the slope of the subgiant branch (see the right-hand
panel of Fig.~11).  Here is an example where it would not be possible to match
the entire C-M diagram, if $(m-M)_V = 14.85$, simply by assuming a different
value of $\alpha_{\rm MLT}$ in the construction of the isochrones: there would
have to be at least two things wrong with the models.  All things considered,
the [Fe/H] value determined by Carretta \& Gratton (1997) for M$\,$3 seems to
be too high. [Even the somewhat higher value obtained by the Lick-Texas
group from high-resolution spectroscopy of bright M$\,$3 giants (i.e., [Fe/H]
$=-1.47$, Kraft et al.~1998) would seem to be too high by 0.1--0.2 dex
according to the present analysis.]

\begin{figure}
\figurenum{12}
\plotfiddle{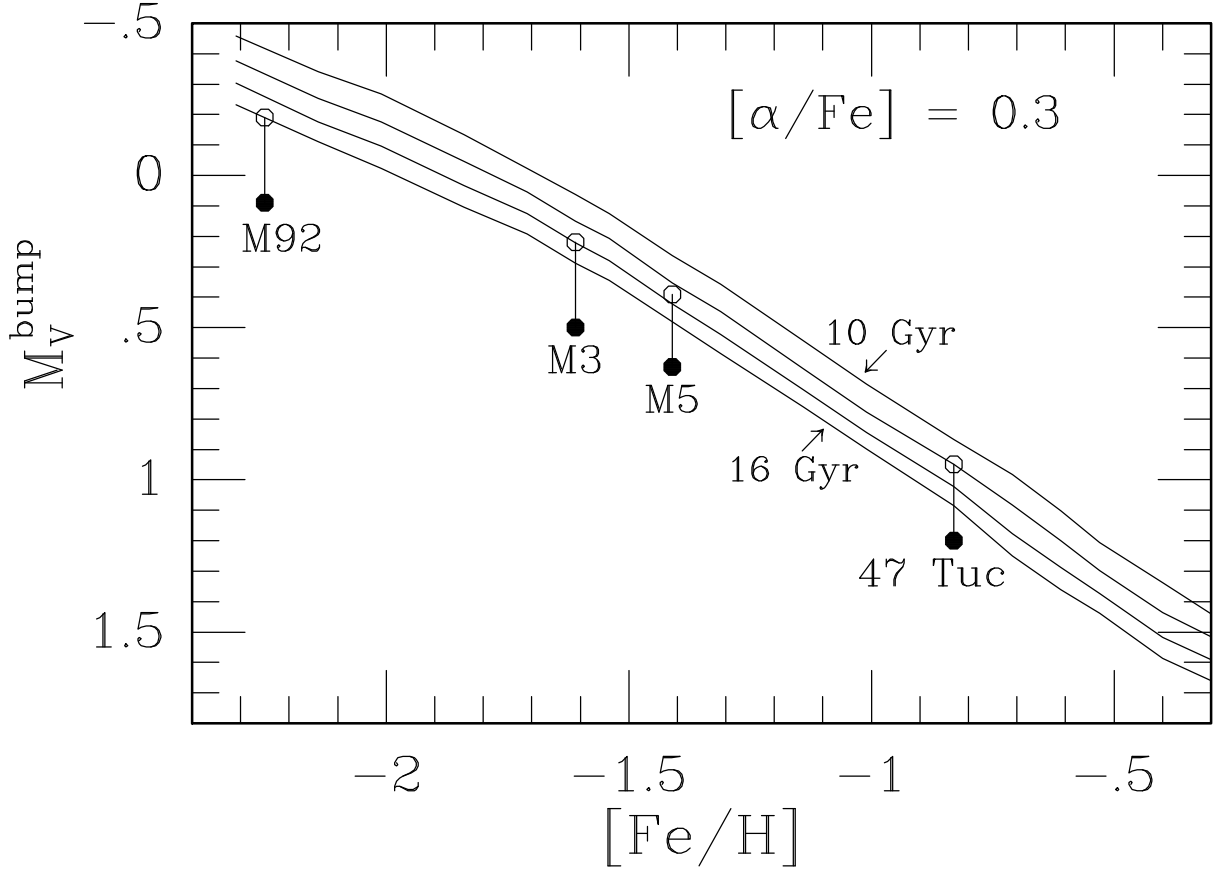}{3.75in}{0}{100}{100}{-306}{-126}
\caption{Plot of the $M_V^{\rm bump}$ versus [Fe/H] data given in
Table~6 for [$\alpha$/Fe] $=0.3$ and ages from 10 to 16 Gyr, in 2 Gyr steps.  If
the ages of M$\,$92, M$\,$3, M$\,$5, and 47 Tuc are 16 Gyr, 14 Gyr, 13 Gyr, and
12 Gyr, respectively (see the text), then the observed magnitudes of the RGB
bumps in these clusters are given by the {\it filled circles}, while the
predicted luminosities of this diagnostic are given by the {\it open circles}.\label{f12}}
\end{figure}

     We conclude this section with Figure~\ref{f12}, which plots the
$M_V^{\rm bump}$ versus [Fe/H] relations given in Table 6 for the [$\alpha$/Fe]
$=0.3$ case.  Superimposed on this diagram are the predicted and observed bump
magnitudes for M$\,$92, M$\,$3, M$\,$5, and 47 Tuc on the assumption of [Fe/H]
values that are within 0.1 dex of those given by Zinn \& West (1984) and which
present relatively few difficulties for our models.  We have also adopted
fairly high ages for these clusters, as indicated by the location of the {\it
open circles} relative to the {\it solid curves}.  This is for the following
reason.  Paper II found that, when zero-age horizontal branch loci are used to
set the globular cluster distance scale, there was a significant dependence of
mean age with metallicity.  To be specific, the mean age varied smoothly from
$\approx 14$ Gyr at [Fe/H]$=-2.3$ to $\approx 11.5$ Gyr at [Fe/H]$=-0.8$.
However, it was also noted that this relation should be displaced to higher
ages if the distances are set instead according to empirical determinations of
the luminosities of RR Lyrae stars from, e.g., Baade-Wesselink and statistical
parallax studies.  As argued therein and in the present investigation, the
short distance scale is also implied by field Population II subdwarfs and
subgiants.

    Thus, in order to satisfy other constraints, we conclude that our models
fail to match the observed luminosities of the RGB bump in clusters by $\approx
0.25$ mag.  This is certainly speculation on our part, but not without
justification.  The location of radiative--convective boundaries in stars is
subject to a lot of uncertain physics and it would not be too surprising that
current models are somewhat lacking in this regard.  Whether or not the scenario
that we have described is realistic may be difficult to confirm or refute until
the uncertainties associated with current distance, $T_{\rm eff}$, and chemical
abundance determinations are significantly reduced.

\section{Concluding Remarks}

The algorithms presented in BV92 to produce isochrones, LFs, and CFs
from grids of evolutionary sequences have been improved with the
implementation of the Akima spline to replace linear interpolation.
Experiments with the interpolating algorithms show that they work well,
over a wide range in metallicity ($+0.12\ge$[Fe/H]$\ge-2.31$) and ages
(from $\sim 1$ Gyr, which is the youngest age that has been examined,
to 20 Gyr), to produce the morphology of isochrones accurately.
Moreover, what we have called the ``isochrone population function",
which literally gives the number of stars at any location along the
corresponding isochrone, has been built into the interpolation
algorithms.

Software (FORTRAN 77) has been developed to implement the algorithms
for grids of evolutionary sequences with $-2.31\le [Fe/H]\le -0.30$.
Each grid of sequences is available with three $\alpha$-element
enhancements: [$\alpha$/Fe]$=0.0$, $+0.3$, and $+0.6$. [Note that the
diffusion of helium and heavy elements has not been treated in the
models.  However, as mentioned at the end of section 4.2 (also see
Grundahl et al.~2000), current observations of Population II stars
appear to be inconsistent with the predictions of diffusive models
(like those recently reported by Salaris, Groenewegen,\& Weiss 2000),
for reasons that have yet to be understood.] The evolutionary grids,
together with the software are available to the astronomical community,
so that users can generate isochrones, IPFs, LFs, and CFs for the
particular set of abundances and ages of interest to them.

The IPF is a particularly powerful tool for stellar astrophysics and
for stellar population studies because it contains all of the
information found in isochrones, LFs, and CFs in a compact form. For
example, magnitudes and color indices for synthetic stellar populations
can easily be generated from the cumulative form of the IPF for any
pair of $B$, $V$, $R$, and $I$, for which empirically constrained
color--$\teff$ relations and bolometric corrections are supplied as
part of the package. (Transformations to the Str\"omgren system,
presently being developed by J.~Clem at the University of Victoria,
will be provided in the near future.  Those involving the $JHK$, and
possibly $U$, bands will be added at a later date.) A side benefit of
the differential IPFs, LFs, and CFs is that they provide a way of fine
tuning the transformation tables that are used to transpose the models
from the $\log L$--$\log \teff$ plane to the observer's plane.

The additional tests of the models that we have carried out to augment
those described in Papers I and II suggest that the computed
temperatures and adopted color--$\teff$\ relations are quite realistic.
The apparent agreement between the predicted and observed properties
of local subdwarfs having well-determined parallaxes ($\sigma_\pi/\pi
\lta 0.1$, as measured by {\it Hipparcos}) is especially encouraging.
Indeed, there should be no difference between the distances which are
derived to e.g., globular star clusters when the observed fiducials are
fitted either to the subdwarf calibrators or to our theoretical isochrones.
However, in practice, the large uncertainty that presently exists in the
GC [Fe/H] scale (see, e.g., \cite{rhs97}) is a severe limitation of the 
main-sequence fitting technique.  Be that as it may, it is our impression
that something close to the Zinn \& West (1984) scale is required to
achieve the best overall consistency between synthetic and observed CMDs
(and between distance estimates based on the subdwarf and RR Lyrae
``standard candles" --- see Paper II).

    Both in zero-point and in slope, and on either the theoretical or
observed planes, our computed giant branches agree well with those
derived observationally.  The models do not reproduce the luminosity of
the RGB bump as well as they should, but this may simply be an
indication that there is some overshooting of the convective envelopes
of red giants into their radiative interiors --- as suggested some time
ago by, e.g., the Padova group (Alongi et al.~1991). In conclusion, we
believe that the present grid of models is particularly well
constrained and it should provide a valuable tool for the
interpretation of stellar data, whether through the fitting of
isochrones, LFs, and CFs or through the construction of integrated
stellar population models whose properties can be compared with those
of distant systems.

\acknowledgements
We thank Raffaele Gratton for providing his estimates of the effective
temperatures for a few field halo stars. The support of Operating
Grants (to P.A.B.~and to D.A.V.) from the Natural Sciences and
Engineering Research Council of Canada is also gratefully
acknowledged.

\newpage

\begin{table}
\tablenum{1}
\center
\newlength{\capla}
\settowidth{\capla}{Table \thetable~~Evolutionary Track (EEP) File Format and Contents}
\parbox{\capla}{\caption{Evolutionary Track (EEP) File Format and Contents}}
\begin{tabular}{l}
\tableline
\tableline
\relax\\[-1.7ex]
{\obeyspaces\tt TRACKS      7}\\
{\obeyspaces\tt [Fe/H]     -0.525}\\
{\obeyspaces\tt [alpha/Fe] +0.3}\\
{\obeyspaces\tt Z          1.010D-02   Z = 6.000D-03 + alpha-element enhancement}\\
{\obeyspaces\tt X          0.742900}\\
{\obeyspaces\tt Y          0.247000}\\
{\obeyspaces\tt ALPHA(mlt) 1.89}\\
{\tt  }\\
{\obeyspaces\tt  Mass  Npts  Match    D(age)    D(log Teff)   Zage      Primary EEPs}\\
{\obeyspaces\tt 1.100   434   194  +4.4901D-04  +2.7616D-04  0.0391  1 36   134198278434}\\
{\obeyspaces\tt   1  0.132689 3.792568 3.9100D-02 0.7407  2.5689242D-02  1.4486261D-03}\\
{\obeyspaces\tt   2  0.133692 3.792632 2.5526D-03 0.7404  2.4341290D-02  1.3111739D-03}\\
{\obeyspaces\tt   3  0.135469 3.792717 8.9341D-03 0.7393  2.0199367D-02  8.8881749D-04}\\
{\obeyspaces\tt   4  0.137098 3.792766 1.3401D-02 0.7376  1.6208193D-02  4.8355037D-04}\\
{\obeyspaces\tt   5  0.138709 3.792789 2.0102D-02 0.7352  1.5496454D-02  4.0982141D-04}\\
{\obeyspaces\tt   .~~~~~~.~~~~~~~~.~~~~~~~~.~~~~~~~~~.~~~~~~~~~.~~~~~~~~~~~~~~.}\\
{\obeyspaces\tt   .~~~~~~.~~~~~~~~.~~~~~~~~.~~~~~~~~~.~~~~~~~~~.~~~~~~~~~~~~~~.}\\
{\obeyspaces\tt   .~~~~~~.~~~~~~~~.~~~~~~~~.~~~~~~~~~.~~~~~~~~~.~~~~~~~~~~~~~~.}\\
{\obeyspaces\tt 434  3.408267 3.530332 1.9975D-04 0.4746  1.9413841D+03 -2.1235817D+02}\\
{\tt  }\\
{\obeyspaces\tt  Mass  Npts  Match    D(age)    D(log Teff)   Zage      Primary EEPs}\\
{\obeyspaces\tt 1.000   432   178  +5.1563D-04  +1.4076D-04  0.0500  1 49   134188268432}\\
{\obeyspaces\tt   1 -0.067340 3.770430 5.0000D-02 0.7407  2.3175805D-02 -5.0453201D-04}\\
{\obeyspaces\tt   2 -0.066481 3.770415 3.3192D-03 0.7404  2.2202865D-02 -4.4192319D-04}\\
{\obeyspaces\tt   3 -0.064795 3.770383 1.1617D-02 0.7394  1.9219138D-02 -2.4991998D-04}\\
{\obeyspaces\tt   4 -0.063101 3.770375 1.7426D-02 0.7379  1.5988973D-02 -6.7321489D-05}\\
{\obeyspaces\tt   5 -0.061384 3.770385 2.6138D-02 0.7356  1.5124787D-02  1.7622827D-04}\\
{\obeyspaces\tt   .~~~~~~.~~~~~~~~.~~~~~~~~.~~~~~~~~~.~~~~~~~~~.~~~~~~~~~~~~~~.}\\
{\obeyspaces\tt   .~~~~~~.~~~~~~~~.~~~~~~~~.~~~~~~~~~.~~~~~~~~~.~~~~~~~~~~~~~~.}\\
{\obeyspaces\tt   .~~~~~~.~~~~~~~~.~~~~~~~~.~~~~~~~~~.~~~~~~~~~.~~~~~~~~~~~~~~.}\\
{\obeyspaces\tt   .~~~~~~.~~~~~~~~.~~~~~~~~.~~~~~~~~~.~~~~~~~~~.~~~~~~~~~~~~~~.}\\
\relax\\[-1.7ex]
\tableline
\end{tabular}
\end{table}

\clearpage

\begin{table}
\tablenum{2}
\center
\newlength{\caplb}
\settowidth{\caplb}{Table \thetable~~Isochrone (ISO) File Format and Contents}
\parbox{\caplb}{\caption{Isochrone (ISO) File Format and Contents}}
\begin{tabular}{l}
\tableline
\tableline
\relax\\[-1.7ex]
{\obeyspaces\tt ISOCHRONES  6}\\
{\obeyspaces\tt [Fe/H]     -0.525}\\
{\obeyspaces\tt [alpha/Fe] +0.3}\\
{\obeyspaces\tt Z          1.010D-02   Z = 6.000D-03 + alpha-element enhancement}\\
{\obeyspaces\tt X          0.742900}\\
{\obeyspaces\tt Y          0.247000}\\
{\obeyspaces\tt ALPHA(mlt) 1.89}\\
{\tt  }\\
{\obeyspaces\tt  Age  Npts}\\
{\obeyspaces\tt  8.0   256}\\
{\obeyspaces\tt   1 -1.333304 3.593967 0.5068505908  0.0000000 4.4097889D-01 4.2796556D-01}\\
{\obeyspaces\tt   2 -1.285759 3.594365 0.5313098653  0.0595653 3.7309103D-01 3.6969990D-01}\\
{\obeyspaces\tt   3 -1.236702 3.596696 0.5532147760  0.1251682 2.9388266D-01 3.0552844D-01}\\
{\obeyspaces\tt   4 -1.187757 3.600882 0.5724189213  0.1992959 2.2815211D-01 2.4122070D-01}\\
{\obeyspaces\tt   5 -1.140842 3.606441 0.5888593464  0.2801004 1.8235053D-01 1.9184230D-01}\\
{\obeyspaces\tt   .~~~~~~.~~~~~~~~.~~~~~~~~~~.~~~~~~~~~~~.~~~~~~~~~~.~~~~~~~~~~~~~.}\\
{\obeyspaces\tt   .~~~~~~.~~~~~~~~.~~~~~~~~~~.~~~~~~~~~~~.~~~~~~~~~~.~~~~~~~~~~~~~.}\\
{\obeyspaces\tt   .~~~~~~.~~~~~~~~.~~~~~~~~~~.~~~~~~~~~~~.~~~~~~~~~~.~~~~~~~~~~~~~.}\\
{\obeyspaces\tt 256  3.408755 3.529098 1.0791668980  7.7891261 2.0186710D-04 1.5983838D-04}\\
{\tt  }\\
{\obeyspaces\tt  Age  Npts}\\
{\obeyspaces\tt 10.0   253}\\
{\obeyspaces\tt   1 -1.339361 3.594616 0.5002053380  0.0000000 4.5027950D-01 4.5082604D-01}\\
{\obeyspaces\tt   2 -1.299791 3.594533 0.5213256448  0.0494690 3.9824028D-01 3.9623085D-01}\\
{\obeyspaces\tt   3 -1.258991 3.595736 0.5405685847  0.1018699 3.3393240D-01 3.3839995D-01}\\
{\obeyspaces\tt   4 -1.217881 3.598282 0.5578690324  0.1592193 2.7157814D-01 2.8065268D-01}\\
{\obeyspaces\tt   5 -1.177527 3.601987 0.5732006971  0.2218015 2.2171010D-01 2.3005911D-01}\\
{\obeyspaces\tt   .~~~~~~.~~~~~~~~.~~~~~~~~~~.~~~~~~~~~~~.~~~~~~~~~~.~~~~~~~~~~~~~.}\\
{\obeyspaces\tt   .~~~~~~.~~~~~~~~.~~~~~~~~~~.~~~~~~~~~~~.~~~~~~~~~~.~~~~~~~~~~~~~.}\\
{\obeyspaces\tt   .~~~~~~.~~~~~~~~.~~~~~~~~~~.~~~~~~~~~~~.~~~~~~~~~~.~~~~~~~~~~~~~.}\\
{\obeyspaces\tt   .~~~~~~.~~~~~~~~.~~~~~~~~~~.~~~~~~~~~~~.~~~~~~~~~~.~~~~~~~~~~~~~.}\\
\relax\\[-1.7ex]
\tableline
\end{tabular}
\end{table}

\clearpage
\begin{table}
\tablenum{3}
\center
\newlength{\caplc}
\settowidth{\caplc}{Table \thetable~~Isochrone Probability Function (IPF) File Format and Contents}
\parbox{\caplc}{\caption{Isochrone Probability Function (IPF) File Format and Contents}}
\scriptsize
\begin{tabular}{l}
\tableline
\tableline
\relax\\[-1.7ex]
{\obeyspaces\tt IPFS        6  I  V-I  x = -0.5 +0.5 +1.5}\\
{\obeyspaces\tt [Fe/H]     -0.525}\\
{\obeyspaces\tt [alpha/Fe] +0.3}\\
{\obeyspaces\tt Z          1.010D-02   Z = 6.000D-03 + alpha-element enhancement}\\
{\obeyspaces\tt X          0.742900}\\
{\obeyspaces\tt Y          0.247000}\\
{\obeyspaces\tt ALPHA(mlt) 1.89}\\
{\tt  }\\
{\obeyspaces\tt  Age  Npts}\\
{\obeyspaces\tt  8.0    99}\\
{\obeyspaces\tt         I     V-I    Mass     Mbol  log Te  log g     d         x = -0.5           x = +0.5           x = +1.5}\\
{\obeyspaces\tt    1  7.262  1.651 0.5418861  7.907 3.5952  4.768   6.400   4.52271  5.00000   4.66645  5.00000   4.79193  5.00000}\\
{\obeyspaces\tt    2  7.101  1.621 0.5669044  7.756 3.5995  4.745   6.200   4.32906  4.97005   4.45372  4.95771   4.55961  4.94257}\\
{\obeyspaces\tt    3  6.980  1.581 0.5833506  7.643 3.6043  4.731   6.000   4.16448  4.94973   4.27571  4.92960   4.36835  4.90503}\\
{\obeyspaces\tt    4  6.891  1.536 0.5956772  7.550 3.6094  4.723   5.800   4.06875  4.93525   4.17060  4.90989   4.25386  4.87902}\\
{\obeyspaces\tt    5  6.811  1.490 0.6063663  7.462 3.6147  4.717   5.600   4.01700  4.92327   4.11100  4.89376   4.18639  4.85793}\\
{\obeyspaces\tt    .~~~~.~~~~~~.~~~~~~~.~~~~~~~~.~~~~~~.~~~~~~.~~~~~~~.~~~~~~~~.~~~~~~~~.~~~~~~~~~.~~~~~~~~.~~~~~~~~~.~~~~~~~~.}\\
{\obeyspaces\tt    .~~~~.~~~~~~.~~~~~~~.~~~~~~~~.~~~~~~.~~~~~~.~~~~~~~.~~~~~~~~.~~~~~~~~.~~~~~~~~~.~~~~~~~~.~~~~~~~~~.~~~~~~~~.}\\
{\obeyspaces\tt    .~~~~.~~~~~~.~~~~~~~.~~~~~~~~.~~~~~~.~~~~~~.~~~~~~~.~~~~~~~~.~~~~~~~~.~~~~~~~~~.~~~~~~~~.~~~~~~~~~.~~~~~~~~.}\\
{\obeyspaces\tt   99 -4.046  3.070 1.0791651 -3.757 3.5297  0.140 -13.187   0.51774 -0.24401   0.36126 -0.40049   0.18615 -0.57560}\\
{\obeyspaces\tt      -4.064  3.091 1.0791669 -3.772 3.5291  0.132 -13.273}\\
{\tt  }\\
{\obeyspaces\tt  Age  Npts}\\
{\obeyspaces\tt 10.0   101}\\
{\obeyspaces\tt         I     V-I    Mass     Mbol  log Te  log g     d         x = -0.5           x = +0.5           x = +1.5}\\
{\obeyspaces\tt    1  7.302  1.653 0.5314385  7.948 3.5950  4.775   6.200   4.61279  5.00000   4.74746  5.00000   4.86529  5.00000}\\
{\obeyspaces\tt    2  7.133  1.626 0.5592527  7.786 3.5986  4.747   6.000   4.41954  4.96284   4.53063  4.94850   4.62527  4.93112}\\
{\obeyspaces\tt    3  6.998  1.590 0.5778846  7.661 3.6034  4.731   5.800   4.22658  4.93724   4.32342  4.91398   4.40341  4.88590}\\
{\obeyspaces\tt    4  6.904  1.545 0.5909741  7.564 3.6083  4.721   5.600   4.12586  4.91999   4.21263  4.89109   4.28255  4.85631}\\
{\obeyspaces\tt    5  6.822  1.500 0.6021440  7.474 3.6135  4.714   5.400   4.06100  4.90581   4.13958  4.87249   4.20130  4.83250}\\
{\obeyspaces\tt    .~~~~.~~~~~~.~~~~~~~.~~~~~~~~.~~~~~~.~~~~~~.~~~~~~~.~~~~~~~~.~~~~~~~~.~~~~~~~~~.~~~~~~~~.~~~~~~~~~.~~~~~~~~.}\\
{\obeyspaces\tt    .~~~~.~~~~~~.~~~~~~~.~~~~~~~~.~~~~~~.~~~~~~.~~~~~~~.~~~~~~~~.~~~~~~~~.~~~~~~~~~.~~~~~~~~.~~~~~~~~~.~~~~~~~~.}\\
{\obeyspaces\tt    .~~~~.~~~~~~.~~~~~~~.~~~~~~~~.~~~~~~.~~~~~~.~~~~~~~.~~~~~~~~.~~~~~~~~.~~~~~~~~~.~~~~~~~~.~~~~~~~~~.~~~~~~~~.}\\
{\obeyspaces\tt    .~~~~.~~~~~~.~~~~~~~.~~~~~~~~.~~~~~~.~~~~~~.~~~~~~~.~~~~~~~~.~~~~~~~~.~~~~~~~~~.~~~~~~~~.~~~~~~~~~.~~~~~~~~.}\\
\relax\\[-1.7ex]
\tableline
\end{tabular}
\end{table}

\clearpage

\begin{table}
\tablenum{4}
\center
\newlength{\capld}
\settowidth{\capld}{Table \thetable~~Luminosity Function (LF) File Format and Contents}
\parbox{\capld}{\caption{Luminosity Function (LF) File Format and Contents}}
%\caption{Luminosity Function (LF) File Format and Contents}
\begin{tabular}{l}
\tableline
\tableline
\relax\\[-1.7ex]
{\obeyspaces\tt LFS         6  I  V-I  x = -0.5 +0.5 +1.5}\\
{\obeyspaces\tt [Fe/H]     -0.525}\\
{\obeyspaces\tt [alpha/Fe] +0.3}\\
{\obeyspaces\tt Z          1.010D-02   Z = 6.000D-03 + alpha-element enhancement}\\
{\obeyspaces\tt X          0.742900}\\
{\obeyspaces\tt Y          0.247000}\\
{\obeyspaces\tt ALPHA(mlt) 1.89}\\
{\tt  }\\
{\obeyspaces\tt  Age  Npts}\\
{\obeyspaces\tt  8.0    57}\\
{\obeyspaces\tt         I         x = -0.5           x = +0.5           x = +1.5}\\
{\obeyspaces\tt    1  7.200   4.54403  5.00000   4.68367  5.00000   4.80591  5.00000}\\
{\obeyspaces\tt    2  7.000   4.46666  4.96849   4.58374  4.95591   4.68324  4.94056}\\
{\obeyspaces\tt    3  6.800   4.43036  4.94023   4.52787  4.91738   4.60780  4.88966}\\
{\obeyspaces\tt    4  6.600   4.32834  4.91252   4.40940  4.88043   4.47284  4.84172}\\
{\obeyspaces\tt    5  6.400   4.27108  4.88928   4.33876  4.85003   4.38882  4.80289}\\
{\obeyspaces\tt    .~~~~~.~~~~~~~.~~~~~~~~.~~~~~~~~.~~~~~~~~~.~~~~~~~~~.~~~~~~~~.}\\
{\obeyspaces\tt    .~~~~~.~~~~~~~.~~~~~~~~.~~~~~~~~.~~~~~~~~~.~~~~~~~~~.~~~~~~~~.}\\
{\obeyspaces\tt    .~~~~~.~~~~~~~.~~~~~~~~.~~~~~~~~.~~~~~~~~~.~~~~~~~~~.~~~~~~~~.}\\
{\obeyspaces\tt   56 -3.800   1.17235  0.73334   1.02008  0.58107   0.85016  0.41115}\\
{\obeyspaces\tt   57 -3.982   1.17301  0.38695   1.02073  0.23467   0.85081  0.06475}\\
{\tt  }\\
{\obeyspaces\tt  Age  Npts}\\
{\obeyspaces\tt 10.0   57}\\
{\obeyspaces\tt         I         x = -0.5           x = +0.5           x = +1.5}\\
{\obeyspaces\tt    1  7.200   4.59146  5.00000   4.71916  5.00000   4.83211  5.00000}\\
{\obeyspaces\tt    2  7.000   4.51210  4.96470   4.61695  4.95194   4.70681  4.93658}\\
{\obeyspaces\tt    3  6.800   4.47533  4.93293   4.56047  4.90980   4.63058  4.88213}\\
{\obeyspaces\tt    4  6.600   4.36671  4.90154   4.43538  4.86909   4.48901  4.83051}\\
{\obeyspaces\tt    5  6.400   4.30264  4.87542   4.35813  4.83585   4.39858  4.78902}\\
{\obeyspaces\tt    .~~~~~.~~~~~~~.~~~~~~~~.~~~~~~~~.~~~~~~~~~.~~~~~~~~~.~~~~~~~~.}\\
{\obeyspaces\tt    .~~~~~.~~~~~~~.~~~~~~~~.~~~~~~~~.~~~~~~~~~.~~~~~~~~~.~~~~~~~~.}\\
{\obeyspaces\tt    .~~~~~.~~~~~~~.~~~~~~~~.~~~~~~~~.~~~~~~~~~.~~~~~~~~~.~~~~~~~~.}\\
{\obeyspaces\tt    .~~~~~.~~~~~~~.~~~~~~~~.~~~~~~~~.~~~~~~~~~.~~~~~~~~~.~~~~~~~~.}\\
\relax\\[-1.7ex]
\tableline
\end{tabular}
\end{table}

\clearpage

\begin{table}
\tablenum{5}
\center
\newlength{\caple}
\settowidth{\caple}{Table \thetable~~Color Function (CF) File Format and Contents}
\parbox{\caple}{\caption{Color Function (CF) File Format and Contents}}
%\caption{Color Function (CF) File Format and Contents}
\begin{tabular}{l}
\tableline
\tableline
\relax\\[-1.7ex]
{\obeyspaces\tt CFS         6  I  V-I  x = -0.5 +0.5 +1.5}\\
{\obeyspaces\tt [Fe/H]     -0.525}\\
{\obeyspaces\tt [alpha/Fe] +0.3}\\
{\obeyspaces\tt Z          1.010D-02   Z = 6.000D-03 + alpha-element enhancement}\\
{\obeyspaces\tt X          0.742900}\\
{\obeyspaces\tt Y          0.247000}\\
{\obeyspaces\tt ALPHA(mlt) 1.89}\\
{\tt  }\\
{\obeyspaces\tt  Age  Npts   Nms}\\
{\obeyspaces\tt  8.0   177    53}\\
{\obeyspaces\tt        V-I         x = -0.5           x = +0.5           x = +1.5}\\
{\obeyspaces\tt    1  1.661   6.15467  5.00000   6.31226  5.00000   6.44939  5.00000}\\
{\obeyspaces\tt    2  1.650   5.48941  4.98464   5.62727  4.97775   5.74506  4.96919}\\
{\obeyspaces\tt    3  1.630   5.16197  4.95595   5.28373  4.93716   5.38481  4.91399}\\
{\obeyspaces\tt    4  1.610   4.96835  4.94176   5.08033  4.91742   5.17185  4.88751}\\
{\obeyspaces\tt    5  1.590   4.85784  4.93242   4.96383  4.90460   5.04942  4.87047}\\
{\obeyspaces\tt    .~~~~~.~~~~~~~.~~~~~~~~.~~~~~~~~.~~~~~~~~~.~~~~~~~~~.~~~~~~~~.}\\
{\obeyspaces\tt    .~~~~~.~~~~~~~.~~~~~~~~.~~~~~~~~.~~~~~~~~~.~~~~~~~~~.~~~~~~~~.}\\
{\obeyspaces\tt    .~~~~~.~~~~~~~.~~~~~~~~.~~~~~~~~.~~~~~~~~~.~~~~~~~~~.~~~~~~~~.}\\
{\obeyspaces\tt   52  0.650   5.54940  4.26728   5.42737  4.12937   5.28500  3.97155}\\
{\obeyspaces\tt   53  0.640   6.23807  4.05760   6.10449  3.90953   5.95048  3.74122}\\
{\obeyspaces\tt   54  0.640   6.23139  4.01804   6.09502  3.86857   5.93823  3.69882}\\
{\obeyspaces\tt   55  0.650   5.20946  3.97522   5.06719  3.82436   4.90450  3.65320}\\
{\obeyspaces\tt    .~~~~~.~~~~~~~.~~~~~~~~.~~~~~~~~.~~~~~~~~~.~~~~~~~~~.~~~~~~~~.}\\
{\obeyspaces\tt    .~~~~~.~~~~~~~.~~~~~~~~.~~~~~~~~.~~~~~~~~~.~~~~~~~~~.~~~~~~~~.}\\
{\obeyspaces\tt    .~~~~~.~~~~~~~.~~~~~~~~.~~~~~~~~.~~~~~~~~~.~~~~~~~~~.~~~~~~~~.}\\
{\obeyspaces\tt  176  3.070   1.11244 -0.39442   0.94847 -0.55839   0.76407 -0.74279}\\
{\obeyspaces\tt  177  3.085   1.12639 -0.84117   0.96242 -1.00514   0.77801 -1.18955}\\
{\tt  }\\
{\obeyspaces\tt  Age  Npts   Nms}\\
{\obeyspaces\tt 10.0   179    50}\\
{\obeyspaces\tt        V-I         x = -0.5           x = +0.5           x = +1.5}\\
{\obeyspaces\tt    1  1.649   5.78583  5.00000   5.92022  5.00000   6.03681  5.00000}\\
{\obeyspaces\tt    2  1.630   5.24754  4.95007   5.35566  4.93044   5.44625  4.90657}\\
{\obeyspaces\tt    3  1.610   5.06810  4.93249   5.16642  4.90668   5.24674  4.87538}\\
{\obeyspaces\tt    4  1.590   4.88341  4.92045   4.97514  4.89059   5.04888  4.85445}\\
{\obeyspaces\tt    5  1.570   4.81476  4.91240   4.90146  4.87991   4.97015  4.84065}\\
{\obeyspaces\tt    .~~~~~.~~~~~~~.~~~~~~~~.~~~~~~~~.~~~~~~~~~.~~~~~~~~~.~~~~~~~~.}\\
{\obeyspaces\tt    .~~~~~.~~~~~~~.~~~~~~~~.~~~~~~~~.~~~~~~~~~.~~~~~~~~~.~~~~~~~~.}\\
{\obeyspaces\tt    .~~~~~.~~~~~~~.~~~~~~~~.~~~~~~~~.~~~~~~~~~.~~~~~~~~~.~~~~~~~~.}\\
{\obeyspaces\tt    .~~~~~.~~~~~~~.~~~~~~~~.~~~~~~~~.~~~~~~~~~.~~~~~~~~~.~~~~~~~~.}\\
\relax\\[-1.7ex]
\tableline
\end{tabular}
\end{table}

\clearpage

\begin{deluxetable}{cccccccccccc}
\tablenum{6}
\tablewidth{500pt}
\tablecaption{Predicted Magnitudes of the RGB Bump as a Function of
Age and Chemical Composition}
\scriptsize
%\tablehead{ \colhead{[Fe/H]} & \colhead{[$\alpha$/Fe]} &
%  \multispan{4}{\hfil $M_V^{\rm bump}$ \hfil} &
%  \colhead{$M_V^{\rm bump}$} & \colhead{$M_V^{\rm bump}$} &
%  \colhead{$M_V^{\rm bump}$} & \colhead{$M_V^{\rm bump}$} & 
%  \colhead{[Fe/H]} & \colhead{[$\alpha$/Fe]} &
%  \multispan{4}{\hfil $M_V^{\rm bump}$ \hfil} \nl&
%  \colhead{$M_V^{\rm bump}$} & \colhead{$M_V^{\rm bump}$} &
%  \colhead{$M_V^{\rm bump}$} & \colhead{$M_V^{\rm bump}$} \nl & &
%  \multispan{4}{\hrulefill} & & & \multispan{4}{\hrulefill} \nl &
\tablehead{ & \colhead{t(Gyr):} & \colhead{10} & \colhead{12} & \colhead{14} &
   \colhead{16} & & \colhead{t(Gyr):} & \colhead{10} &\colhead{12} &
   \colhead{14} & \colhead{16} \nl \noalign{\vskip-3pt} & &
   \multispan{4}{\hrulefill} & & & \multispan{4}{\hrulefill} \nl
   \colhead{[Fe/H]} & \colhead{[$\alpha$/Fe]} &
   \multispan{4}{\hfil $M_V^{\rm bump}$ \hfil} & \colhead{[Fe/H]} &
   \colhead{[$\alpha$/Fe]} & \multispan{4}{\hfil $M_V^{\rm bump}$ \hfil} }
\startdata
$-2.31$ & 0.00 & $-0.59$ & $-0.50$ & $-0.43$ & $-0.35$ &
 $-2.14$ & 0.00 & $-0.48$ & $-0.39$ & $-0.31$ & $-0.23$ \nl
        & 0.30 & $-0.46$ & $-0.38$ & $-0.30$ & $-0.23$ &
         & 0.30 & $-0.34$ & $-0.25$ & $-0.18$ & $-0.11$ \nl
        & 0.60 & $-0.31$ & $-0.23$ & $-0.16$ & $-0.10$ &
         & 0.60 & $-0.17$ & $-0.09$ & $-0.03$ & $+0.03$ \nl
\noalign{\vskip3pt}
$-2.01$ & 0.00 & $-0.41$ & $-0.31$ & $-0.22$ & $-0.14$ &
 $-1.84$ & 0.00 & $-0.31$ & $-0.21$ & $-0.13$ & $-0.06$ \nl
        & 0.30 & $-0.27$ & $-0.18$ & $-0.10$ & $-0.02$ &
         & 0.30 & $-0.13$ & $-0.05$ & $+0.03$ & $+0.10$ \nl
        & 0.60 & $-0.08$ & $+0.01$ & $+0.09$ & $+0.15$ &
         & 0.60 & $+0.07$ & $+0.16$ & $+0.23$ & $+0.29$ \nl
\noalign{\vskip3pt}
$-1.71$ & 0.00 & $-0.19$ & $-0.10$ & $-0.03$ & $+0.04$ &
 $-1.61$ & 0.00 & $-0.12$ & $-0.03$ & $+0.07$ & $+0.13$ \nl
        & 0.30 & $-0.02$ & $+0.05$ & $+0.13$ & $+0.19$ &
         & 0.30 & $+0.06$ & $+0.15$ & $+0.22$ & $+0.29$ \nl
        & 0.60 & $+0.16$ & $+0.25$ & $+0.34$ & $+0.42$ &
         & 0.60 & $+0.28$ & $+0.36$ & $+0.43$ & $+0.50$ \nl
\noalign{\vskip3pt}
$-1.53$ & 0.00 & $-0.05$ & $+0.05$ & $+0.12$ & $+0.19$ &
 $-1.41$ & 0.00 & $+0.08$ & $+0.17$ & $+0.25$ & $+0.31$ \nl
        & 0.30 & $+0.13$ & $+0.21$ & $+0.28$ & $+0.34$ &
         & 0.30 & $+0.26$ & $+0.35$ & $+0.42$ & $+0.48$ \nl
        & 0.60 & $+0.34$ & $+0.43$ & $+0.51$ & $+0.58$ &
         & 0.60 & $+0.45$ & $+0.54$ & $+0.63$ & $+0.71$ \nl
\noalign{\vskip3pt}
$-1.31$ & 0.00 & $+0.15$ & $+0.25$ & $+0.34$ & $+0.41$ &
 $-1.14$ & 0.00 & $+0.31$ & $+0.41$ & $+0.51$ & $+0.59$ \nl
        & 0.30 & $+0.36$ & $+0.45$ & $+0.52$ & $+0.59$ &
         & 0.30 & $+0.54$ & $+0.64$ & $+0.71$ & $+0.76$ \nl
        & 0.60 & $+0.58$ & $+0.66$ & $+0.74$ & $+0.81$ &
         & 0.60 & $+0.77$ & $+0.86$ & $+0.94$ & $+1.00$ \nl
\noalign{\vskip3pt}
$-1.01$ & 0.00 & $+0.45$ & $+0.54$ & $+0.63$ & $+0.71$ &
 $-0.83$ & 0.00 & $+0.67$ & $+0.77$ & $+0.84$ & $+0.91$ \nl
        & 0.30 & $+0.69$ & $+0.78$ & $+0.84$ & $+0.90$ &
         & 0.30 & $+0.87$ & $+0.95$ & $+1.02$ & $+1.09$ \nl
        & 0.60 & $+0.91$ & $+1.01$ & $+1.09$ & $+1.16$ &
         & 0.60 & $+1.11$ & $+1.20$ & $+1.28$ & $+1.34$ \nl
\noalign{\vskip3pt}
$-0.71$ & 0.00 & $+0.79$ & $+0.88$ & $+0.95$ & $+1.02$ &
 $-0.61$ & 0.00 & $+0.91$ & $+1.01$ & $+1.08$ & $+1.14$ \nl
        & 0.30 & $+0.98$ & $+1.08$ & $+1.18$ & $+1.25$ &
         & 0.30 & $+1.10$ & $+1.20$ & $+1.29$ & $+1.36$ \nl
        & 0.60 & $+1.24$ & $+1.34$ & $+1.42$ & $+1.49$ &
         & 0.60 & $+1.34$ & $+1.43$ & $+1.50$ & $+1.58$ \nl
\noalign{\vskip3pt}
$-0.53$ & 0.00 & $+0.99$ & $+1.09$ & $+1.18$ & $+1.25$ &
 $-0.40$ & 0.00 & $+1.13$ & $+1.23$ & $+1.32$ & $+1.39$ \nl
        & 0.30 & $+1.21$ & $+1.30$ & $+1.37$ & $+1.44$ &
         & 0.30 & $+1.34$ & $+1.44$ & $+1.52$ & $+1.59$ \nl
        & 0.60 & $+1.42$ & $+1.52$ & $+1.60$ & $+1.67$ &
         & 0.60 & $+1.54$ & $+1.65$ & $+1.73$ & $+1.80$ \nl
\noalign{\vskip3pt}
$-0.30$ & 0.00 & $+1.26$ & $+1.34$ & $+1.40$ & $+1.48$ & & & & & & \nl
        & 0.30 & $+1.44$ & $+1.52$ & $+1.59$ & $+1.66$ & & & & & & \nl
        & 0.60 & $+1.65$ & $+1.75$ & $+1.81$ & $+1.87$ & & & & & & \nl
\enddata
\end{deluxetable}

\clearpage

\begin{deluxetable}{ccccccc}
\tablenum{7}
\tablewidth{380pt}
\tablecaption{Comparison of the Observed and Predicted RGB Bump
Magnitudes in M$\,$92}
\tablehead{
\colhead{[Fe/H]} & \colhead{[$\alpha$/Fe]} & \colhead{$(m-M)_V$} &
\colhead{t(Gyr)} & \colhead{$M_{V,\,{\rm obs}}^{\rm bump}$} &
\colhead{$M_{V,\,{\rm theory}}^{\rm bump}$} & \colhead{$\Delta M_V^{\rm bump}$}}
\startdata
$-2.31$ & 0.30 & 14.90 & 12 & $-0.25$ & $-0.38$ & $+0.13$ \nl
        &      & 14.70 & 14 & $-0.05$ & $-0.30$ & $+0.25$ \nl
        &      & 14.57 & 16 & $+0.08$ & $-0.23$ & $+0.31$ \nl
\noalign{\vskip5pt}
$-2.14$ & 0.30 & 14.85 & 12 & $-0.20$ & $-0.25$ & $+0.05$ \nl
        &      & 14.67 & 14 & $-0.02$ & $-0.18$ & $+0.16$ \nl
        &      & 14.54 & 16 & $+0.11$ & $-0.11$ & $+0.22$ \nl
\noalign{\vskip2pt}
\enddata
\end{deluxetable}

\clearpage

\clearpage

\clearpage

\clearpage

\clearpage

\clearpage

\clearpage

\clearpage

\clearpage

\clearpage

\clearpage

\clearpage

\end{document}